\documentclass[11pt]{article}

\usepackage[final]{acl}

\usepackage{times}
\usepackage{latexsym}
\usepackage{colortbl}
\usepackage[T1]{fontenc}

\usepackage[utf8]{inputenc}

\usepackage{microtype}

\usepackage{inconsolata}

\usepackage{booktabs}
\usepackage{caption}      

\usepackage{array}
\usepackage{ragged2e}
\usepackage{enumitem}
\usepackage{pifont}
\usepackage{tikz}
\usepackage{xcolor}
\usepackage{graphicx}
\usepackage{xspace}
\usepackage{stfloats} 

\usepackage[most]{tcolorbox}
\usepackage{setspace} 
\definecolor{customblue}{RGB}{25, 18, 180}

\usepackage{hyperref}
\hypersetup{
  colorlinks   = true,    
  breaklinks   = true,
}

\usepackage[nameinlink, capitalise]{cleveref}       %

\usepackage{cancel}
\usepackage{bigdelim, rotating}  %

\newcommand{\eg}{\emph{e.g}.\xspace}
\newcommand{\ie}{\emph{i.e}.\xspace}

\usetikzlibrary{tikzmark}
\makeatletter
\newcommand*\myfontsize{%
\@setfontsize\myfontsize{6.7}{8}%
}
\makeatother

\definecolor{cadmiumgreen}{rgb}{0.0, 0.42, 0.24}

\definecolor{myred}{rgb}{0.7, 0.3, 0.0}
\definecolor{myblue}{rgb}{0.2, 0.3, 0.6}

\newcommand{\mymidrule}[1]{%
    \noalign{\vskip -0.2\aboverulesep} 
    \cmidrule[\heavyrulewidth]{#1}  
    \noalign{\vskip -0.2\belowrulesep} 
}

\newenvironment{packeditemize}{
\begin{list}{$\bullet$}{
\setlength{\labelwidth}{6pt}
\setlength{\itemsep}{0pt}
\setlength{\leftmargin}{\labelwidth}
\addtolength{\leftmargin}{\labelsep}
\setlength{\parindent}{0pt}
\setlength{\listparindent}{\parindent}
\setlength{\parsep}{0pt}
\setlength{\topsep}{3pt}}}{\end{list}}

\renewcommand{\texttt}[1]{%
  \begingroup
  \ttfamily
  \begingroup\lccode`~=`/\lowercase{\endgroup\def~}{/\discretionary{}{}{}}%
  \begingroup\lccode`~=`[\lowercase{\endgroup\def~}{[\discretionary{}{}{}}%
  \begingroup\lccode`~=`.\lowercase{\endgroup\def~}{.\discretionary{}{}{}}%
  \catcode`/=\active\catcode`[=\active\catcode`.=\active
  \scantokens{#1\noexpand}%
  \endgroup
}

\newcommand{\llm}[1]{\texttt{#1}}
\newcommand{\dataset}[1]{\textbf{\textsc{#1}}}

\newcolumntype{L}[1]{>{\raggedright\arraybackslash}p{#1}}
\newcolumntype{C}[1]{>{\centering\arraybackslash}p{#1}}
\newcolumntype{R}[1]{>{\raggedleft\arraybackslash}p{#1}}
\newcolumntype{M}[1]{>{\centering\arraybackslash}m{#1}}
\newcolumntype{P}[1]{>{\raggedright\arraybackslash}m{#1}}

\NewEnviron{promptgroup}{%
  \begin{minipage}{\columnwidth}
    \BODY
  \end{minipage}
}

\newcounter{sharedbox}                   
\crefname  {sharedbox}{Prompt}{Prompts}  
\Crefname  {sharedbox}{Prompt}{Prompts}  

\newtcolorbox{promptbox}[2][]{%
  float,                    
  floatplacement=!ht,      
  width=\linewidth,         
  colback=white,            
  colframe=black,           
  coltitle=black,           
  colbacktitle=white,       
  boxrule=1.2pt,
  left=5pt, right=5pt, top=5pt, bottom=5pt,
  before upper={\setstretch{0.8}},  
  fonttitle=\sffamily\bfseries\small,
  title={%
    \refstepcounter{sharedbox}
    \label{#1}
    \fontsize{9.7}{7}\selectfont Prompt \thesharedbox: #2%
  },%
}

\newtcolorbox{promptboxc}[2][]{%
  float*=!ht,                 
  width=\textwidth,
  colback=white,
  colframe=black,
  coltitle=black,
  colbacktitle=white,
  boxrule=1.2pt,
  left=5pt, right=5pt, top=5pt, bottom=5pt,
  before upper={\setstretch{0.8}}, 
  fonttitle=\sffamily\bfseries\small,
  title={%
    \refstepcounter{sharedbox}%
    \label{#1}%
    \fontsize{9.7}{7}\selectfont Prompt \thesharedbox: #2%
  },%
}
\newtcolorbox[auto counter]{examplebox}[2][]{
  float,
  float=htbp,  
  width=\linewidth,
  colback=white,
  title={\fontsize{9.7}{7}\selectfont Example \thetcbcounter: #2},
  coltitle=black,
  left=5pt,
  right=5pt,
  top=5pt,
  bottom=5pt,
  fonttitle=\sffamily\bfseries\small,
  boxrule=1.2pt,
  label={#1},
  colframe=black,
  colbacktitle=white,
  before upper={\setstretch{0.9}},
  before={\par\vspace*{0pt}},
  after={\par\vspace*{0pt}},
}

\crefname{figure}{Figure}{Figures}
\Crefname{figure}{Figure}{Figures}
\crefname{table}{Table}{Tables}
\Crefname{table}{Table}{Tables}
\crefname{equation}{Eq.}{Eqs.}
\Crefname{equation}{Eq.}{Eqs.}
\crefname{appendix}{Appendix}{Appendices}
\Crefname{appendix}{Appendix}{Appendices}

\makeatletter
\crefname{tcb@cnt@promptbox}{prompt}{prompts}    
\Crefname{tcb@cnt@promptbox}{Prompt}{Prompts}    
\makeatother

\makeatletter
\crefname{tcb@cnt@examplebox}{example}{examples}    
\Crefname{tcb@cnt@examplebox}{Example}{Examples}    
\makeatother

\definecolor{lightblue}{RGB}{220,235,250}

\tcbset{
  takeawaysbox/.style={
    title=Takeaways,
    colback=lightblue!80,
    colframe=black,
    fonttitle=\bfseries\small,
    coltitle=white,
    colbacktitle=black,
    enhanced,
    attach boxed title to top left={xshift=2.5mm,yshift=-2.5mm},
    boxed title style={rounded corners, size=small, colframe=black, colback=black},
    width=\linewidth,
    arc=3.5mm
  }
}

\usepackage{arydshln} 

\usepackage{hyperref}
\usepackage{url}
\usepackage{wrapfig}   
\usepackage{booktabs}  
\usepackage{tikz}
\usepackage{pgfcalendar}
\usepackage{etoolbox}
\usepackage{xparse}
\usepackage{graphicx}    
\usepackage{subcaption}
\usepackage{wasysym}
\usetikzlibrary{positioning, arrows.meta}
\usepackage{minitoc}
\usepackage{makecell}
\usepackage{xcolor, color, soul}

\usepackage{amsmath,amsfonts,bm}

\def\eqref#1{equation~\ref{#1}}

\def\1{\bm{1}}

\DeclareMathAlphabet{\mathsfit}{\encodingdefault}{\sfdefault}{m}{sl}
\SetMathAlphabet{\mathsfit}{bold}{\encodingdefault}{\sfdefault}{bx}{n}

\newcommand{\ourbench}{\dataset{IFEval++}\xspace}
\newcommand{\reliableatk}{\texttt{reliable@k}\xspace}
\newcommand{\reliableat}[1]{\texttt{reliable@{#1}}\xspace}

\title{Revisiting the Reliability of Language Models in Instruction-Following}

\author{
Jianshuo Dong$^{1}$, Yutong Zhang$^{1}$, Yan Liu$^{2}$, Zhenyu Zhong$^{2}$,\\
\textbf{Tao Wei$^{2}$, Chao Zhang$^{1}$, Han Qiu$^{1}$\thanks{Corresponding author.}} \\
$^{1}$Tsinghua University, China. $^{2}$Ant Group, China.\\
\texttt{dongjs23@mails.tsinghua.edu.cn, qiuhan@tsinghua.edu.cn}
}

\begin{document}
\maketitle

\begin{abstract}
Advanced LLMs have achieved near-ceiling instruction-following accuracy on benchmarks such as \dataset{IFEval}.
However, these impressive scores do not necessarily translate to reliable services in real-world use, where users often vary their phrasing, contextual framing, and task formulations.
In this paper, we study \emph{nuance-oriented reliability}: whether models exhibit consistent competence across \textit{cousin prompts} that convey analogous user intents but with subtle nuances.
To quantify this, we introduce a new metric, \reliableatk, and develop an automated pipeline that generates high-quality cousin prompts via data augmentation.
Building upon this, we construct \ourbench\ for systematic evaluation.
Across 20 proprietary and 26 open-source LLMs, we find that current models exhibit substantial insufficiency in nuance-oriented reliability---their performance can drop by up to 61.8\% with nuanced prompt modifications.
What's more, we characterize it and explore three potential improvement recipes.
Our findings highlight nuance-oriented reliability as a crucial yet underexplored next step toward more dependable and trustworthy LLM behavior.
Our code and benchmark are accessible: \url{https://github.com/jianshuod/IFEval-pp}.
\end{abstract}

\section{Introduction}

Large Language Models (LLMs) have demonstrated remarkable abilities to follow natural language instructions~\citep{ouyang2022instructgpt}.
The instruction-following ability is foundational for enabling faithful user interaction~\citep{zhou2023instruction-ifeval-verification-instruction-following-evaluation-25-types,lior2025wildifeval,zhang2024cfbench-comprehensive-constraints-following-benchmark-for-llms}, reliable agent behavior~\citep{zhang2025which-multi-agent-system-failure-attribution}, and reduced harmful outcomes~\citep{ruan2024identifying-tool-emu-simulate-tools-to-test-agent-safety}.
Evaluating this ability is essential for building reliable, trustworthy AI systems. 
Consequently, numerous benchmarks have been proposed to measure instruction-following performance across diverse task types and constraint categories (see~\Cref{appx:related-works} for a survey of 36 benchmarks).

\begin{figure}[t]
    \centering
    \includegraphics[width=0.5\textwidth]{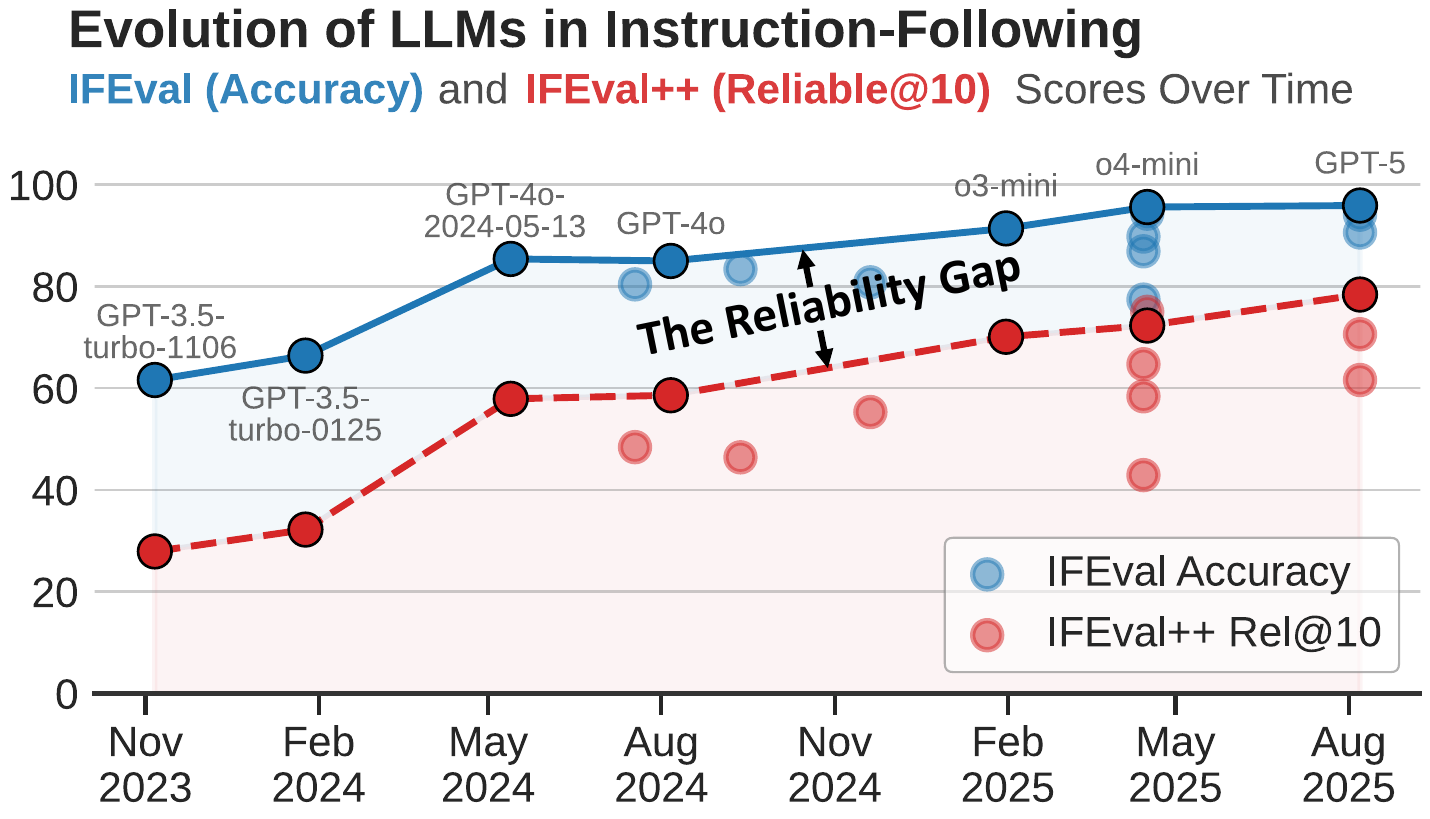}
    \vspace{-1.5em}
    \caption{\textbf{Evolution of OpenAI's Models in Instruction-Following Abilities Across Generations}.}
    \label{fig:illustration}
    \vspace{-1em}
\end{figure}

As illustrated in~\Cref{fig:illustration}, the instruction-following ability of LLMs has improved rapidly across generations.
Recent LLMs achieve near-ceiling performance on widely adopted benchmarks like \dataset{IFEval}~\citep{zhou2023instruction-ifeval-verification-instruction-following-evaluation-25-types}, with \llm{GPT-5} reaching an accuracy as high as 95.9\%.
However, growing evidence reveals that model performance can be highly sensitive to prompt wording~\citep{sclar2024quantifying-how-i-learned-to-start-worrying-about-prompt-formatting-classification-tasks,mizrahi2024state-multi-prompt-llm-evaluation,cao2024worst-prompt-performance-of-llms}, and that overfitting to benchmark test cases is possible~\citep{zhang2024a-benchmark-over-fitting}. 
Thus, questions remain about how well benchmark accuracy translates to real-world reliability. 

To illustrate this concern, consider two LLMs that both achieve perfect benchmark accuracy.
When faced with user prompts expressing similar intent but differing subtly in phrasing, contextual framing, or task instantiation, one model may generalize appropriately while the other fails.
Current benchmarks, which primarily emphasize task and constraint diversity, cannot capture this crucial \textbf{nuance-oriented dimension of reliability}. 
This motivates our central question: \textit{Are current LLMs reliable when handling ``cousin prompts'' that convey similar user intents but vary in nuanced ways?}

\begin{figure*}
    \centering
    \includegraphics[width=1\linewidth]{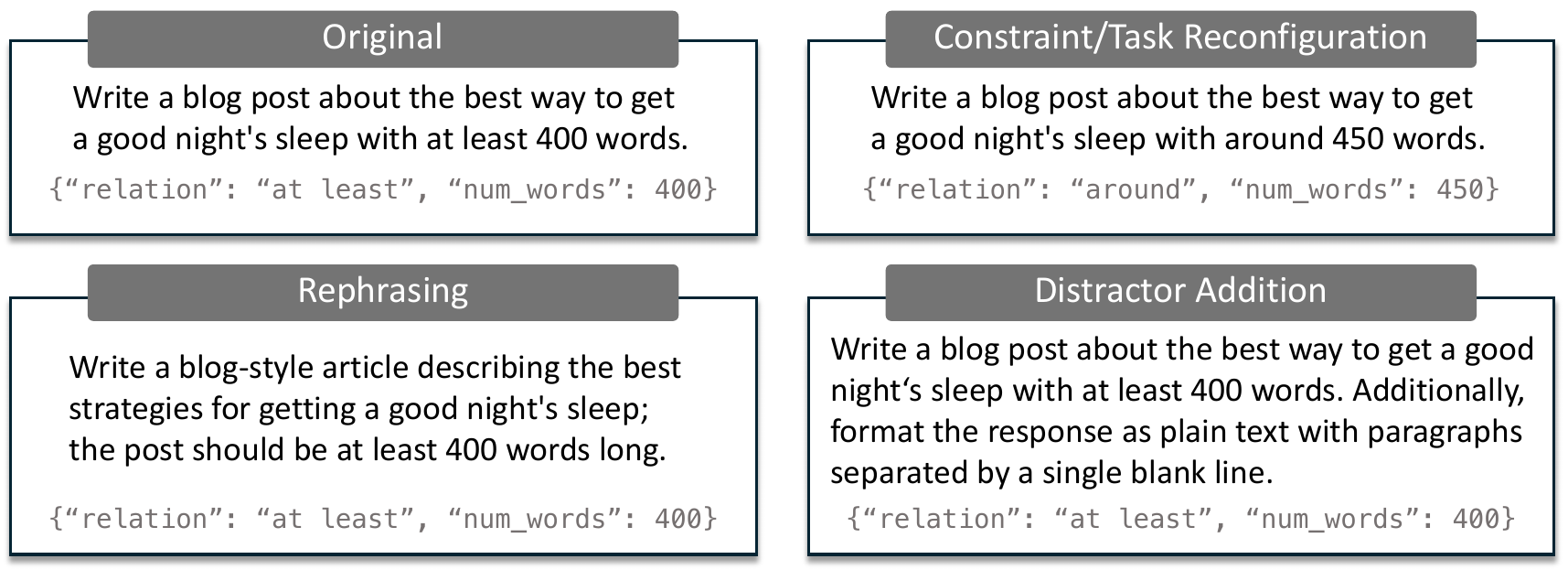}
    \vspace{-2em}
    \caption{\textbf{Examples of Cousin Prompts}. The corresponding evaluation configurations are provided.}
    \label{fig:cousin-inst-examples}
    \vspace{-1em}
\end{figure*}

To operationalize this, we propose a new metric, \reliableatk, which quantifies whether an LLM can simultaneously solve a set of \textit{cousin prompts}.
We further design an automated pipeline for generating high-quality cousin prompts using three data augmentation strategies: rephrasing, distractor addition, and task/constraint reconfiguration.
This is coupled with a code-assisted validity checker. 
Building on \dataset{IFEval}~\citep{zhou2023instruction-ifeval-verification-instruction-following-evaluation-25-types}, we leverage the pipeline to construct an extended benchmark, \ourbench\footnote{
We name it to show respect for the renowned \dataset{IFEval} benchmark, as we position \ourbench as a successor of \dataset{IFEval}, similar to how C++ relates to C.
}, which consists of 541 test cases, each comprising 10 cousin prompts.
This provides a testbed for assessing nuance-oriented reliability of LLMs in a systematic way.

Using \ourbench, we comprehensively evaluate 20 proprietary and 26 open-source LLMs.
Our analysis reveals that current LLMs often exhibit considerable inconsistent competence across cousin prompts.
The relative drop from accuracy on \dataset{IFEval} to \reliableat{10} on \ourbench can be as large as 61.8\% for \llm{Qwen3-0.6B} and 54.7\% for \llm{GPT-3.5-turbo-1106}, while even the most reliable model, \llm{GPT-5}, experiences a decrease of 18.3\%.
Notably, nuance-oriented reliability emerges as \textbf{a second-order property}: for instance, while \llm{Gemma-3-IT-27B} ranks 17th in accuracy on \dataset{IFEval}, it rises to 7th when considering \reliableat{10} on \ourbench.
Furthermore, we empirically characterize how models' nuance-oriented reliability evolves with chronological development, model scale, reasoning capability, and augmentation type.

Observing the insufficiency of current LLMs in nuance-oriented reliability, we investigate three potential pathways for improvement: prediction-based methods, training-based methods, and test-time scaling.
Among these, parallel test-time scaling through rejection sampling proves most effective, enabling a relatively weak model such as \llm{Qwen3-4B} to surpass even the strongest open-source model, \llm{LLaMA-3.3-70B-Instruct}.

We contend that improving this \emph{nuance-oriented reliability} is a crucial next step toward dependable, trustworthy LLM behaviors.
In summary, our contributions primarily lie in:

\noindent $\bullet$ We investigate the nuance-oriented reliability of LLMs in instruction-following and propose a new metric, \reliableatk, to quantify it.

\noindent $\bullet$ We develop an automated pipeline for generating cousin prompts, and build \ourbench, a benchmark that enables actionable evaluation of nuance-oriented reliability.

\noindent $\bullet$ Using \ourbench, we conduct comprehensive experiments across a wide range of models and explore three distinct pathways to enhance reliability.

\section{Preliminaries}\label{subsec:preliminaries}

\subsection{Problem Statement}
\label{subsec:problem-statement}
Let $F$ be an auto-regressive language model that takes a prompt $x$ and produces a response $y = F(x)$.
Following prior work~\citep{zhou2023instruction-ifeval-verification-instruction-following-evaluation-25-types}, instruction-following ability is typically evaluated by checking whether $y$ satisfies a set of constraints $C = \{c_1, c_2, \ldots, c_m\}$ specified in the prompt as natural language instructions.
Please refer to~\Cref{appx:related-works} for related works.

\begin{figure*}[t]
    \centering
    \includegraphics[width=\linewidth]{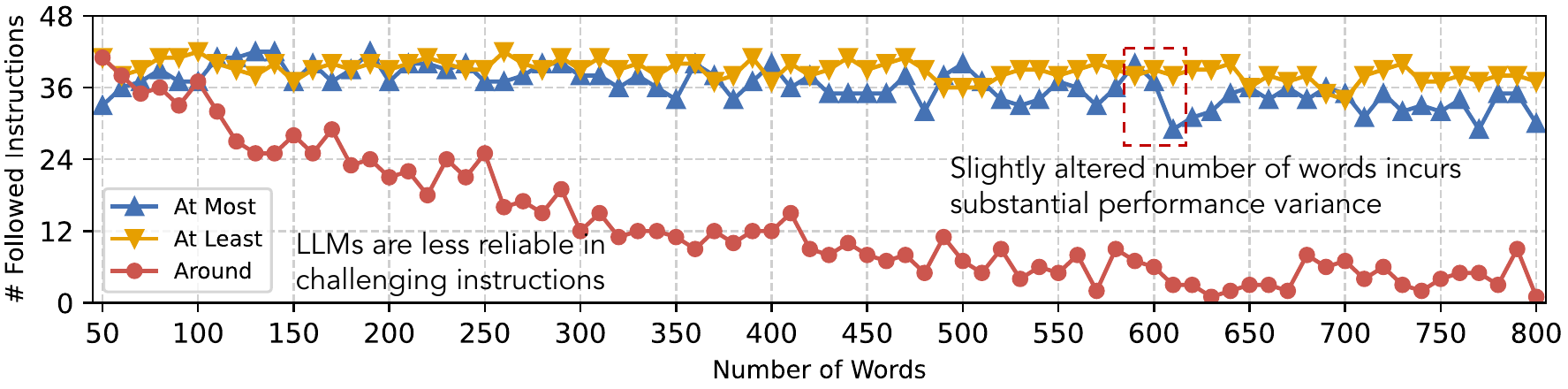}
    \vspace{-2em}
    \caption{\textbf{Results of Requesting Varying Number of Response Words}. Experiments with \llm{Qwen3-8B}.}
    \label{fig:poc-results}
    \vspace{-1em}
\end{figure*}

\noindent \textbf{A Pilot Experiment}.
To ground our discussion, we begin with a pilot experiment focusing on test cases involving word-length constraints on responses. 
From \dataset{IFEval}, we extract 48 suitable prompts characterized by a relation (\textit{at least}, \textit{at most}, or \textit{around}) and a numerical value \textit{num\_word}, as exemplified in~\Cref{fig:cousin-inst-examples}.
We then prompt \llm{GPT-5-mini} to convert the prompt in each test case into a template, leaving the relation and value configurable.
Using a length interval of 10, ranging from 50 to 800, we program to instantiate the templates, yielding $3 \times 76 \times 48 = 10,944$ prompts.
The evaluation configurations are adjusted accordingly.
As shown in~\Cref{fig:poc-results}, LLM performance is highly sensitive to the requested word count.
A minor modification, \eg, from ``at most 600'' to ``at most 610'', can incur many test cases to fail.
This sensitivity is especially pronounced under the more challenging \textit{around} constraint.
This confirms that LLMs may not be reliable across prompts with subtle nuances.
We ask: Shouldn't a reliable LLM tackle them all?

\noindent \textbf{Two Reliability Dimensions in Instruction-Following}.
To situate our research, we review 36 benchmarks designed to evaluate instruction-following performance (see~\Cref{appx:if-benchmarks}).
Most of these works focus on \textit{benchmark comprehensiveness}---seeking evaluations that span diverse task types, domains, and constraint categories.
This focus aligns with the view that a reliable assistant should handle diverse user demands across scenarios.
However, our pilot study suggests that comprehensiveness alone is insufficient to capture the full notion of reliability.
In real-world use, users may phrase identical requirements in multiple ways or express similar requirements with nuanced differences (see~\Cref{fig:cousin-inst-examples} for examples).
A truly reliable LLM should consistently produce correct outputs across these prompts.
We therefore propose a complementary perspective: \textbf{nuance-oriented reliability}, which measures the model's stability when responding to ``cousin prompts'' that convey analogous user intents and differ only in subtle linguistic or semantic nuances.

\subsection{Scope of This Work}
\noindent \textbf{New Metric: \reliableatk}.
To capture it, we introduce a more challenging metric, \reliableatk, designed to provide an actionable evaluation of LLMs' nuance-oriented reliability. 
Specifically, \reliableatk quantifies how consistently an LLM can handle a set of closely related cousin prompts simultaneously.
Formally, for $k$ cousin prompts with model outputs $\{y_1, y_2, \ldots, y_k\}$, we define
\[
\text{\reliableatk} = \mathbb{I}\left( \sum_{j=1}^{k} \text{is\_passed}(y_j) = k \right),
\]
where $\text{is\_passed}(y_j)$ indicates whether the model's response to the $j$-th cousin prompt satisfies the evaluation criterion.
Cousin prompts can be obtained by augmenting the original prompts in existing benchmarks, which we detail in~\Cref{sec:dataset-curation}.
The parameter $k$ and the choice of cousin prompts together control the rigor of the nuance-oriented reliability evaluation, rendering \reliableatk scalable (see~\Cref{appx:relatk-scalability} for experimental evidence).
Notably, when $k=1$, \reliableat{1} reduces to the standard accuracy metric, since only the original prompts are included.
It is worth noting that the \reliableatk metric is orthogonal yet complementary to the \texttt{pass\^{}k} metric~\citep{yao2025taubench-tool-calling-agent}, which runs the same prompt with $k$ independent runs.
We discuss their difference in~\Cref{appx:related-work-reliability}.

\noindent \textbf{Data Source \& Constraint Types}.
We choose \dataset{IFEval}~\citep{zhou2023instruction-ifeval-verification-instruction-following-evaluation-25-types} as our main testbed, since SOTA LLMs have already achieved strong performance on it (\eg, OpenAI's \llm{GPT-5} reaches an accuracy of 95.9\% as measured in~\Cref{tab:main-results}), making it a meaningful setting for discussing reliability and the challenging \reliableatk metric.
This benchmark focuses on single-turn tasks with 25 types of code-verifiable format constraints, such as ``\textit{Answer using at most \{num\_words\} words}.''
Reliably tackling such \textbf{\emph{simple yet foundational}} instructions is a necessary condition for addressing more complex, multi-turn, and multi-modal instruction-following settings.
See~\Cref{appx:constraint-type} for the 25 constraints.

\section{Curating Cousin Prompts At Scale}
\label{sec:dataset-curation}

We introduce three data augmentation methods for producing cousin prompts (\Cref{subsec:data-augmentation}), our checker for ensuring augmented test case validity (\Cref{subsec:checker-design}), and the operational procedures for finalizing the \ourbench benchmark (\Cref{subsec:benchmark-finalization}).

\begin{figure*}[t]
    \centering
    \includegraphics[width=1\textwidth]{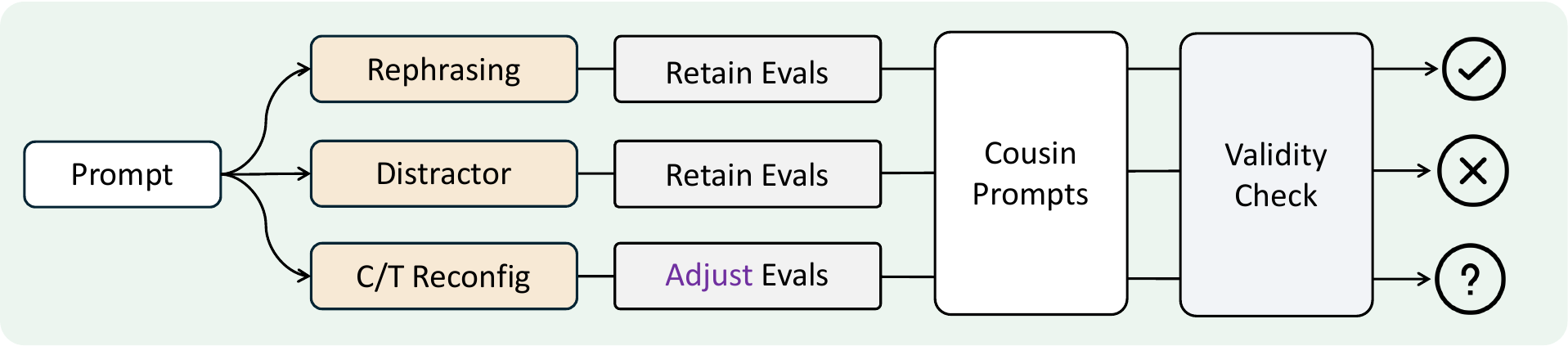}
    \vspace{-1.5em}
    \caption{\textbf{Conceptual Illustration of the Augmentation-Filtering Data Curation Pipeline}.}
    \label{fig:data-curation-pipeline}
    \vspace{-1em}
\end{figure*}

\subsection{Data Augmentation Methods}
\label{subsec:data-augmentation}

Data augmentation for cousin prompts primarily involves revisions on two components: rewriting the prompts and adjusting the evaluation configurations, if necessary.
In this work, we mainly account for three types of augmentation methods and employ LLM-based augmentation.
See examples of the augmented cousin prompts in~\Cref{fig:cousin-inst-examples}.

\textbf{Type 1. Rephrasing.}
This reflects the scenario where different users may articulate the same task in varying ways.
The augmenter rephrases each of the 541 prompts from \dataset{IFEval} into alternative wordings while preserving the core task and constraint semantics (see~\Cref{prompt:rephrasing}).
For rephrasing, the corresponding evaluation configurations remain unchanged, since the rephrased prompts still request responses that satisfy the same requirements as the original prompts.

\textbf{Type 2. Distractor Addition.}
The augmenter appends each prompt with a distractor constraint that does not interfere with the satisfaction of the original constraints (see~\Cref{prompt:distractor}).
The augmenter is instructed to ensure that each distractor constraint remains compatible with the original prompt.
In this setting, responses continue to be evaluated against the original constraints, with the added constraint functioning purely as a distractor, in contrast to prior work on multi-constraint composition~\citep{jiang2024followbench-multi-constraints-composition-and-considering-decoding-methods,wen2024benchmarking-complexbench-multiple-constraints-composition}. 
This setup examines whether LLMs can integrate additional constraints in a plug-and-play manner.

\textbf{Type 3. Constraint/Task Reconfiguration.}
In this setting, we aim to examine whether LLMs can reliably follow instructions when prompts are presented in varied scenarios. 
To this end, we craft cousin prompts that share the same requirement types but differ in subtle ways.
For test cases with at least one configurable constraint (\eg, \textit{num\_word}), we instruct the augmenter to modify the constraint and update the evaluation configurations accordingly (see~\Cref{prompt:constraint-alteration}). 
This mirrors the pilot experiment introduced in~\Cref{subsec:problem-statement}.
For cases with only non-configurable constraints, such as requiring responses in \textit{JSON} format, we instead configure the task to indirectly impose the same requirement in a different context (\Cref{prompt:task-alteration}). 
In the case of task reconfiguration, the evaluation configurations remain unchanged.

\subsection{Code-Assisted Validity Check}
\label{subsec:checker-design}

LLM-based augmentation enables us to rapidly acquire cousin prompts, but risks introducing flawed test cases. 
To mitigate this issue, we develop an automated validity checker.

The checker is designed with an emphasis on high recall: its primary objective is to flag any potentially erroneous cases, even at the cost of over-detection. 
This strategy allows us to curate a high-quality evaluation set by filtering candidates against the checker. 
Inspired by~\citet{chen2023program-of-thoughts-prompting}, our central design is to embed the implementations of evaluation functions directly into the checker's prompt, clearly specifying how the evaluation operates (see~\Cref{prompt:sanity-checker}).
Empirically, this approach enables more deterministic and robust verification.
The checker is implemented in a unified way to monitor all three types of augmented test cases.

\noindent \textbf{Validating the Checker}.
Following principles in~\Cref{appx:checker-principles}, we iteratively refined the checker prompt on $3 \times 300 = 900$ test cases, where the augmenter was deliberately instructed to inject errors.
This is an aggressive modeling of flawed prompts that the checker may encounter.
Using \llm{GPT-5-mini} (medium reasoning effort), the checker achieved a recall of 99.7\% (898/900).

\definecolor{reasonblue}{HTML}{E8EEF7}
\definecolor{chatbrown}{HTML}{F9F2EB}
\definecolor{gptgreen}{HTML}{E4F6EB}
\definecolor{textreasonblue}{HTML}{2563EB}
\definecolor{textchatbrown}{HTML}{F9F2EB}

\newcommand{\reasonblue}{\cellcolor{reasonblue}}
\newcommand{\chatbrown}{\cellcolor{chatbrown}}
\newcommand{\gptgreen}{\rowcolor{gptgreen}}

\newcommand{\onlyatk}[1]{\texttt{rel@{#1}}\xspace}
\newcommand{\datacolwidth}{0.8cm}

\newcommand{\datawithdec}[2]{#1 & -#2\%}

\begin{table*}[!t]
    \centering
\setlength{\tabcolsep}{9pt}
    \caption{\textbf{Nuance-Oriented Reliability of Representative LLMs}. We use \onlyatk{k} for \reliableat{k} and ``C/T'' for constraints or tasks. We report \reliableatk across \ourbench subsets by augmentation method. Models are ranked by \reliableat{10} on \ourbench (with the rankings of \dataset{IFEval} accuracy for reference). Reasoning models are in \colorbox{reasonblue}{\textit{blue}}, non-reasoning in \colorbox{chatbrown}{\textit{beige}}. Best and second-best scores are in \textbf{bold} and \underline{underline}.}
    \label{tab:main-results}
    \resizebox{0.98\textwidth}{!}{
    \begin{tabular}{p{4.3cm}C{\datacolwidth}C{\datacolwidth}C{\datacolwidth}C{\datacolwidth}C{\datacolwidth}C{\datacolwidth}C{\datacolwidth}C{\datacolwidth}C{\datacolwidth}C{1.2cm}}
    \toprule
    \multirow{2}{*}{\textbf{Model}} & \multicolumn{2}{c}{\textbf{\dataset{IFEval}}} &  
    \multicolumn{2}{c}{\textbf{+Rephrasing}}  &  
    \multicolumn{2}{c}{\textbf{+Distractor}} &  
    \multicolumn{2}{c}{\textbf{+C/T Reconfig}}  & \multicolumn{2}{c}{\textbf{\ourbench}} \\  
    \cmidrule(lr){2-3}\cmidrule(lr){4-5}\cmidrule(lr){6-7}\cmidrule(lr){8-9}\cmidrule(lr){10-11}\vspace{-2em}
         & Acc & Rank & \onlyatk{2} & \onlyatk{4} & \onlyatk{2} & \onlyatk{4} & \onlyatk{2} & \onlyatk{4} & \onlyatk{10} & Rel $\Delta$ \\
            \mymidrule{1-11} \rowcolor[gray]{0.9}
\multicolumn{11}{@{}c@{}}{\textbf{\textit{Open-Source LLMs}}} \\
    \mymidrule{1-11}
    \chatbrown LLaMA-3.3-70B-Instruct     & \textbf{92.1}  &  1  &  \textbf{89.6} & \textbf{88.9} & \textbf{88.0} & \textbf{80.0} & \textbf{89.6} & \textbf{80.6} & \textbf{71.0} &  \textbf{-22.9\%} \\
    \reasonblue DeepSeek-V3.1   & 90.6  & 5  & 86.0 & 81.3 & 85.6 & 77.3 & \underline{86.5} & 76.0 & \datawithdec{\underline{63.4}}{30.0} \\
    \reasonblue Qwen3-Next-80B-A3B-TK   &  \underline{91.3}  & 2  & 86.1 & 80.8 & 84.8 & 76.5 & 86.3 & \underline{76.9} & \datawithdec{63.2}{30.8} \\
    \reasonblue Kimi-K2-0905   & \underline{91.3}  & 2  & \underline{86.9} & \underline{81.5} & 85.8 & \underline{78.0} & 85.2 & 76.0 & \datawithdec{63.0}{31.0} \\
    \chatbrown Qwen3-Next-80B-A3B-IT   & 90.9  & 4  & 84.6 & 80.4 & \underline{86.5} & 77.3 & \underline{86.5} & 75.8 & \datawithdec{62.3}{31.5} \\
    \reasonblue GLM-4.5     & 86.9   &  10 & 82.6 & 78.2 & 81.1 & 72.5 & 83.2 & 74.9 & \datawithdec{61.7}{28.9} \\
    \chatbrown Gemma-3-IT-27B     & 84.3   & 17  & 80.4 & 77.8 & 79.1 & 73.4 & 80.4 & 70.2 & 61.6 & \underline{-27.0\%} \\
    \reasonblue Qwen3-14B     & 88.4 & 6  & 85.4 & \underline{81.5} & 83.2 & 73.4 & 83.2 & 72.5 & \datawithdec{61.2}{30.8} \\
    \chatbrown Gemma-3-IT-12B     & 84.3  &  17 & 81.1 & 78.2 & 79.7 & 73.0 & 79.3 & 69.5 & \datawithdec{59.5}{29.4} \\
    \reasonblue Qwen3-8B  & 87.6 & 7 & 83.4 & 79.5 & 81.7 & 72.8 & 81.3 & 69.9  & \datawithdec{58.8}{32.9} \\
    \chatbrown LLaMA-3.1-70B-Instruct    & 85.6  & 15  & 82.3 & 79.1 & 79.5 & 71.0 & 81.0 & 69.1 & \datawithdec{57.1}{33.3} \\
    \reasonblue Qwen3-30B-A3B    & 87.6 & 7 & 80.6 & 75.6 & 81.5 & 70.6 & 82.6 & 71.2 & \datawithdec{57.1}{34.8} \\
    \reasonblue GLM-4.5-Air     & 87.4   & 9  & 82.6 & 76.3 & 81.1 & 70.8 & 81.5 & 70.8 & \datawithdec{56.7}{35.1} \\
    \reasonblue Qwen3-235B-A22B  & 86.3 & 12 & 81.0 & 73.8 & 78.6 & 69.7 & 81.3 & 70.1 & \datawithdec{56.6}{34.5} \\
    \reasonblue GPT-oss-120b    &  86.0 & 14 & 81.9 & 75.4 & 80.0 & 72.6 & 77.3 & 67.1 & \datawithdec{56.4}{34.4} \\
    \chatbrown Qwen2.5-72B-Instruct   & 86.7 &  11  & 82.6 & 76.3 & 77.6 & 68.2 & 80.0 & 69.1 & \datawithdec{55.6}{35.9} \\
    \reasonblue Qwen3-32B     &  86.3 & 12 & 81.5 & 75.8 & 81.9 & 72.3 & 81.7 & 69.7 & \datawithdec{55.3}{36.0} \\
    \reasonblue DeepSeek-R1   & 83.4 & 20   & 77.1 & 70.6 & 77.8  & 69.1 & 75.8 & 65.4 & \datawithdec{53.0}{36.4} \\
    \reasonblue Qwen3-4B  & 85.2 & 16 & 79.9 & 74.5 & 78.7 & 67.8 & 79.1 & 65.4 & \datawithdec{52.1}{38.8} \\
    \chatbrown Qwen2.5-32B-Instruct   & 82.4 &  21  & 77.6 & 71.2 & 72.5 & 62.8 & 75.4 & 64.1 & \datawithdec{50.3}{39.0} \\
    \reasonblue GPT-oss-20b  & 83.9 & 19 & 76.5 & 68.2 & 76.9 & 66.2 & 73.2 & 61.9 & \datawithdec{46.8}{44.3} \\
    \chatbrown Qwen2.5-14B-Instruct   & 78.2 &  22  & 72.6 & 67.1 & 69.5 & 59.0 & 69.7 & 57.3 & \datawithdec{44.7}{42.8} \\
    \chatbrown LLaMA-3.1-8B-Instruct    & 75.8 &  23  & 70.8 & 64.7 & 65.8 & 57.1 & 67.1 & 53.8 & \datawithdec{41.4}{45.4} \\
    \chatbrown Qwen2.5-7B-Instruct   & 73.0 &  24  & 66.2 & 60.1 & 62.5 & 50.5 & 63.8 & 50.6 & \datawithdec{34.8}{52.3} \\ 
    \reasonblue Qwen3-1.7B  & 71.5 & 25 & 63.4 & 55.1 & 62.3 & 49.4 & 61.6 & 46.2 & \datawithdec{34.0}{52.5} \\
    \reasonblue Qwen3-0.6B     &  58.0 & 26 & 47.5 & 41.4 & 46.0 & 33.3 & 46.6 & 34.2 & \datawithdec{22.2}{61.8} \\
        \mymidrule{1-11} \rowcolor[gray]{0.9}
\multicolumn{11}{@{}c@{}}{\textbf{\textit{Proprietary LLMs}}} \\
    \mymidrule{1-11}
    \reasonblue  GPT-5   & \textbf{95.9} &  1  & \textbf{93.3} & \textbf{90.4} & \textbf{93.0} & \textbf{86.5} & \textbf{93.7} & \textbf{88.7} & \textbf{78.4} & \textbf{-18.3\%} \\
    \reasonblue o3   &  94.3 & 3  & \underline{93.5} & \underline{89.1} & \underline{92.1} & \underline{84.5} & 91.7 & \underline{85.0} &  \underline{75.0} & \underline{-21.3\%} \\
    \reasonblue  o4-mini & \underline{95.6} & 2   & 91.9 & 87.1 & 90.4 & 81.7 & \underline{91.9} & 84.8 &  \datawithdec{72.3}{24.4} \\
    \reasonblue  Gemini-2.5-pro   & 93.3 &  5  & 89.6 & 87.4 & 88.4 & 82.6 & 91.3 & 83.0 & 71.9 & -23.0\% \\
    \reasonblue  GPT-5-mini & 94.1 &  4  & 90.6 & 85.2 & 89.3 & 83.0 & 89.1 & 82.8 & \datawithdec{70.6}{25.0} \\
    \reasonblue o3-mini   &  91.4 & 6 & 89.5 & 86.0 & 88.7 & 81.5 & 88.5 & 80.6 & \datawithdec{70.1}{25.2} \\
    \reasonblue Gemini-2.5-flash  & 90.4 &  8  & 87.6  & 84.3 & 85.6 & 78.4 & 86.7 & 78.2 & \datawithdec{66.4}{26.5} \\
    \chatbrown Gemini-2.0-flash  & 90.6 &  7  & 87.8  & 83.7 & 84.4 & 78.0 & 85.6 & 77.3 & \datawithdec{65.8}{27.3} \\
    \chatbrown GPT-4.1 & 89.8 &  9  & 86.3 & 83.0 & 83.9 & 76.3 & 84.3 & 76.9 & \datawithdec{64.7}{28.0} \\
    \reasonblue Gemini-2.5-flash-lite  & 88.0 &  10  & 83.9 & 81.5 & 83.2 & 75.8 & 83.0 & 73.2 & \datawithdec{62.7}{28.8} \\
    \reasonblue  GPT-5-nano   & 90.6 &  7  & 85.0 & 81.3 & 85.8 & 77.3 & 83.9 & 71.5 & \datawithdec{61.6}{32.0} \\
    \chatbrown GPT-4o      &  85.0 & 13 & 80.2 & 76.3 & 79.7 & 71.3 & 79.3 & 69.1 & \datawithdec{58.6}{31.1} \\
    \chatbrown GPT-4.1-mini  & 86.9 &  11  & 83.2 & 78.9 & 79.9 & 71.7 & 80.6 & 70.2 & \datawithdec{58.4}{32.8} \\
    \chatbrown GPT-4o-2024-05-13     &  85.4 & 12 & 81.3 & 78.0 & 79.1 & 71.7 & 79.1 & 66.7 & \datawithdec{57.9}{32.3} \\
    \chatbrown GPT-4o-2024-11-20   &  80.8 & 15 & 77.6 & 73.9 & 75.4 & 68.8 & 75.2 & 64.1 & \datawithdec{55.3}{31.6} \\    
    \chatbrown GPT-4o-mini     &  80.4 & 16 & 75.8 & 71.7 & 72.1 & 64.1 & 72.8 & 60.3 & \datawithdec{48.4}{39.8} \\
    \reasonblue o1-mini    &  83.4 & 14 & 76.2 & 66.9 & 74.7 & 65.2 & 73.8 & 59.3 & \datawithdec{46.4}{44.3} \\
    \chatbrown GPT-4.1-nano & 77.4 &  17  & 72.5 & 67.8 & 70.1 & 60.6 & 68.8 & 55.6 & \datawithdec{42.9}{44.6} \\
    \chatbrown GPT-3.5-turbo-0125     &  66.4 & 18 & 57.9 & 50.8 & 55.5 & 44.7 & 55.3 & 42.1 & \datawithdec{32.2}{51.5} \\
    \chatbrown GPT-3.5-turbo-1106     &  61.6 & 19 & 53.0 & 48.4 & 50.5 & 41.0 & 51.8 & 37.8 & \datawithdec{27.9}{54.7} \\
    
    \bottomrule
    \end{tabular}
    }
\end{table*}

To further assess robustness, we evaluated it on $3 \times 1,000 = 3,000$ additional flawed cases, maintaining a recall of 99.9\% (2,997/3,000); the few misses were due to unavoidable hallucinations~\citep{sun2025detection-hallucination-of-lrms}. 
The checker also achieved 90.0\% precision on 541 unrevised prompts\footnote{
Many flagged cases were indeed flawed or ambiguous. Necessary corrections are detailed in~\Cref{appx:revise-test-cases}, and the revised dataset is used for subsequent experiments.
}.
Overall, the checker rigorously detects invalid prompts with minimal valid loss.
Moreover, since the augmenter is also explicitly guided to avoid introducing flaws, the two-layer mechanism minimizes the risk of flaws arising from the augmentation process.

\subsection{Synthesizing the \ourbench Benchmark}
\label{subsec:benchmark-finalization}

In this section, we synthesize \ourbench: 
To balance cost and coverage, each original test case is expanded into nine replicates, with three from each augmentation method.
Following the augmentation methodology introduced earlier, we automatically produce cousin prompts and verify them using the code-assisted validity checker to ensure integrity.
This process is iterated until a sufficient number of valid cousin prompts are collected.

The \ourbench benchmark comprises 541 test cases, each containing one original prompt and nine cousin prompts.
With \ourbench, we can characterize the nuance-oriented dimension of LLMs' reliability, which cannot be revealed by evaluating only standalone prompts.
Accordingly, we report \reliableat{1} (\ie, accuracy on \dataset{IFEval}), \reliableat{10} on \ourbench, and \reliableat{4} and \reliableat{2} on the subsets of \ourbench corresponding to each augmentation type.

\section{Experiments}

\subsection{Experimental Setup}
\label{subsec:exp-setup}

\noindent \textbf{Models}.
In this work, we include a diverse set of 46 models to facilitate comparative analyses across four key dimensions: (a) differences in model scale, (b) chronologically different generations of models, (c) contrasts between open-source and proprietary models, and (d) distinctions between reasoning and non-reasoning models. 
Additional model details are provided in~\Cref{appx:model-details}.

\noindent \textbf{Generation Strategies}.
We use official default system messages when provided in chat template (\eg, for \llm{Qwen-2.5} models); otherwise, we leave the system message empty to elicit the model's default behavior, as suggested by models like \llm{DeepSeek-R1}\footnote{\url{https://huggingface.co/deepseek-ai/DeepSeek-R1\#usage-recommendations}}.
We further validate this choice in~\Cref{appx:system-prompt-impact}.
We employ greedy decoding by default, which is suitable for reliability evaluation and provides a reproducibility guarantee.
Additionally, we enable reasoning mode by default if supported.

\noindent \textbf{Evaluation Methods}.
Following the practices of \llm{DeepSeek-V3}~\citep{liu2024deepseek-v3}, \llm{Qwen3}~\citep{qwen2025qwen3technicalreport}, and \llm{Kimi K2}~\citep{team2025kimi-k2}, we use the \textit{prompt strict} evaluation mode of \dataset{IFEval}.
This requires LLMs to satisfy all requirements specified in the prompts.
Based on this success signal, we compute the \reliableatk metric.

\subsection{Main Results with \ourbench}\label{subsec:main-results}

We answer the initial question:  
\textbf{Current LLMs do not guarantee nuance-oriented reliability.}  

As shown in~\Cref{tab:main-results}, the relative drop from accuracy on \dataset{IFEval} to \reliableat{10} on \ourbench is substantial, reaching up to 61.8\% for \llm{Qwen3-0.6B} and 54.7\% for \llm{GPT-3.5-turbo-1106}.  
Even the most reliable model, \llm{GPT-5}, suffers an 18.3\% decline when confronted with prompts containing subtle variations.  
Importantly, we observe that higher accuracy on \dataset{IFEval} does not necessarily translate into higher \reliableat{10} on \ourbench.
For example, while \llm{Gemma-3-IT-27B} ranks only 17th on \dataset{IFEval}, it rises to 7th place on \ourbench.
This highlights that \textbf{nuance-oriented reliability is a second-order property beyond accuracy}, one that comprehensiveness-oriented benchmarks fail to capture.
We provide the category-wise reliability in~\Cref{appx:category-wise-reliability}.
Besides, we arrive at several empirical findings:

\noindent $\bullet$ \textbf{Chronological Development}:  
As expected, more recent LLMs tend to perform better, especially when developed by the same vendor.  
This underscores the importance of the underlying training methodology, as highlighted by the stark difference in reliability between \llm{LLaMA-3.3-70B-Instruct} (\reliableat{10}: 71.0) and its predecessor, \llm{LLaMA-3.1-70B-Instruct} (\reliableat{10}: 57.1).  

\noindent $\bullet$ \textbf{Model Scale}:  
In general, larger-scale models tend to outperform smaller ones. 
An exception is observed within the \llm{Qwen3} family, where \llm{Qwen3-14B} demonstrates lower accuracy on \dataset{IFEval} but achieves a higher \reliableat{10} score compared to the larger \llm{Qwen3-32B}. 
This indicates that surpassing a certain parameter size is sufficient to attain high reliability. 
Meanwhile, performance is not determined by scale alone: 
Training data quality and methodology also play crucial roles. 
Notably, models with comparable parameter counts but developed by different vendors can show substantial differences in reliability, as exemplified by \llm{Gemma-3-IT-27B} and \llm{Qwen2.5-32B-Instruct}.

\noindent $\bullet$ \textbf{Reasoning}:  
Reasoning models demonstrate increasingly stronger nuance-oriented reliability.  
However, \llm{LLaMA-3.3-70B-Instruct}, though not a reasoning model, still achieves the highest ranking among open-source models.
This suggests reasoning itself may not be a prerequisite for achieving high reliability.
This motivates us to examine the specific contribution of reasoning to reliability, which we investigate in~\Cref{subsec:test-time-scaling}.

\noindent $\bullet$ \textbf{Augmentation Type}. 
Models exhibit varying robustness to the three types of cousin prompts.
Most models can effectively handle rephrased cousin prompts, which share the greatest similarity with the original ones.
This result is expected, as rephrased prompts differ only in wording (understanding the instructions) while requesting the same underlying response (executing the instructions).
In contrast, cousin prompts that include additional distractor constraints or C/T reconfiguration introduce greater complexity.
Their corresponding \reliableat{4} scores are typically lower than those of rephrased prompts.
We hypothesize that, although distractor constraints are designed to remain compatible with the original instructions, and C/T reconfiguration introduces only slight variations in the underlying requirements, both inevitably increase the difficulty of response planning and execution~\citep{dong2025emergent-planning-pre-generation-probing}.

\begin{table}[t]
    \centering
    \caption{\textbf{AUROC Scores of Prediction Methods}.}
    \label{tab:following-prediction}
    \resizebox{0.46\textwidth}{!}{
    \begin{tabular}{M{0.25\columnwidth} M{0.28\columnwidth} M{0.28\columnwidth}}
        \toprule
        \textbf{Method} & \llm{Qwen3-8B} & \llm{Qwen2.5-7B} \\
        \midrule
        Verb. Conf.   & 0.549 & 0.518 \\
        Perplexity    & 0.497 & 0.529 \\
        Probing       & 0.757 & 0.759 \\
        \bottomrule
    \end{tabular}
    }
    \vspace{-1em}
\end{table}

\section{How to Improve Reliability?}

While LLMs are not yet reliable, we explore three possible directions for improving them.

\subsection{Predicting the Selective Following}
\label{sec:instruction-following-prediction}

If we can predict how the LLM reacts to different prompts, we can proactively approach reliable LLM services by prompt selection.
This case study goes in a similar vein to~\citet{stolfo2024improving-instruction-following-via-activation-steering,cao2024worst-prompt-performance-of-llms,heo2025do-llms-know-internally-when-they-follow-instructions,heo2025do-instruction-following-uncertainty-post-response}, but our focus is to predict the instruction-following prior to generation and across various cousin prompts.
We experiment with \llm{Qwen3-8B} and \llm{Qwen2.5-7B-Instruct}.
For each model, we randomly sample 1,200 of them as the validation set, with 600 followed and 600 unfollowed.
We evaluate three prediction methods, with performance presented in~\Cref{tab:following-prediction}.

\noindent $\bullet$ \textbf{Verbalized Confidence}.
We instruct the LLM to articulate its confidence in successfully following one given prompt, on a scale from 0 to 9.
Yet, the self-reported confidence yields performance close to random guessing.
This outcome can be attributed to the LLMs' general tendency toward overconfidence, aligning with~\citet{xiong2024can-llms-express-their-uncertainty}.

\noindent $\bullet$ \textbf{Prompt Perplexity}.
We compute the perplexity of LLMs on each prompt and use it for predicting instruction-following.
Yet, the close-to-chance AUROC reveals that low-perplexity prompts that are most familiar to the model do not necessarily correspond to successful instruction-following.

\noindent $\bullet$ \textbf{Probing Hidden States}. 
Causal LMs may plan their responses before generation, and the relevant signals can be probed from their hidden states~\citep{dong2025emergent-planning-pre-generation-probing}. 
Specifically, we train a linear logistic regression probe, which takes the hidden states of the last input token as input.
To train the probe, we additionally collect a non-overlapping set of 200 followed and 200 unfollowed prompts.
Compared with the previous two methods, probing provides a more effective prediction.
The best validation performance is achieved when probing the 17th and the 27th layer on two models, respectively.
Although probing shows potential in this direction, the probes are far from being a reliable predictor.

\begin{figure}[t]
    \centering
    \includegraphics[width=0.47\textwidth]{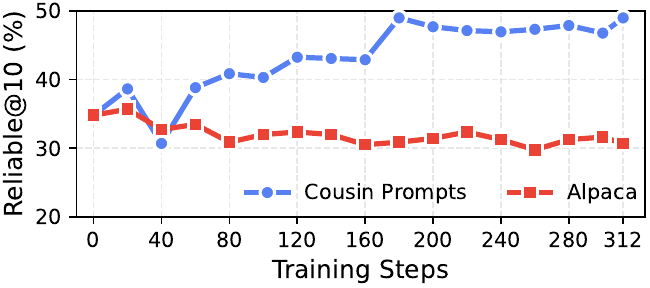}
    \vspace{-0.5em}
    \caption{\textbf{Training on Datasets Affects Reliability Performance on \ourbench in Varying Fashion}.}
    \label{fig:training-effects}
    \vspace{-1em}
\end{figure}

\subsection{Training-Based Method}
\label{subsec:training-based}

We adopt supervised fine-tuning (SFT) as our training method and evaluate two dataset configurations, which represent the two extremes of prompt relatedness.
(1) We use the general-domain instruction-following dataset, \dataset{alpaca}~\citep{alpaca}, whose prompts are largely unrelated to \ourbench.
(2) We curate an additional set of prompts, also augmented from \dataset{IFEval} but decontaminated from~\ourbench prompts via exact matching.
Then, we collect responses via rejection sampling against \llm{LLaMA-3.3-70B-Instruct}.
We fine-tune \llm{Qwen-2.5-7B-Instruct} for 312 steps on each dataset.
Further details on data curation and training are provided in~\Cref{appx:training-details}.

\Cref{fig:training-effects} reports the reliability performance on \ourbench across training steps. 
The two datasets lead to markedly different outcomes. 
Models fine-tuned on \dataset{alpaca} exhibit a slightly declining performance. 
In contrast, training on the curated \textit{cousin prompts} steadily improves reliability, with performance climbing above 45\% after 200 steps.
This reaffirms reliability as a second-order property from the training perspective.
It benefits more from targeted fine-tuning on semantically adjacent samples rather than from the sheer scale of training data.
This observation resonates with recent explorations on rewriting training data~\citep{fujii2025rewriting-pre-training-data-boosts-llm-performance-in-math-and-code,team2025kimi-k2}.
In~\Cref{appx:vs-dpo}, we additionally explore how different training techniques (SFT vs. DPO~\citep{rafailov2024dpo}) affect the nuance-oriented reliability.

\begin{figure}[t]
    \centering
    \includegraphics[width=0.47\textwidth]{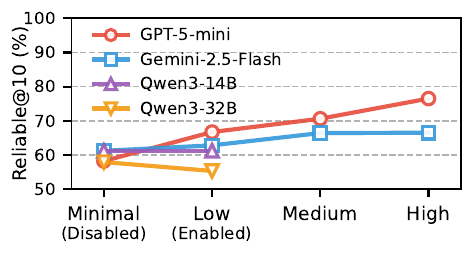}
    \vspace{-1em}
    \caption{\textbf{Impact of Reasoning Effort on Nuance-Oriented Reliability of LLMs}.}
    \label{fig:reasoning-effort}
    \vspace{-1em}
\end{figure}

\subsection{Test-Time Scaling}
\label{subsec:test-time-scaling}

We explore two forms of scaling up test-time compute~\citep{snell2024scaling-test-time-compute-optimality}: sequential and parallel.

\noindent $\bullet$ \textbf{Sequential (Reasoning Effort)}.
Reasoning models inherently capture thinking patterns such as self-correction~\citep{guo2025deepseek-r1}. 
Scaling up reasoning effort can be viewed as a sequential analogue to scaling up test-time compute.
We include four models that support adaptive reasoning budget control: \llm{GPT-5-mini}, \llm{Gemini-2.5-Flash}, \llm{Qwen3-14B}, and \llm{Qwen3-32B}.
As shown in~\Cref{fig:reasoning-effort}, increasing reasoning effort usually improves reliability, particularly for \llm{GPT-5-mini}.
Manual inspection of their intrinsic CoT reveals that, through reasoning, models can infer user intent and adapt responses accordingly.
This explicit reasoning process enhances robustness against subtle prompt variations.
However, inappropriate reasoning patterns may reduce reliability~\citep{chen2024not-tencent-llm-overthinking,dong2025towards-lrm-cog-habits}, as evidenced by the negative effects observed in \llm{Qwen3-32B}.

\begin{figure}[t]
    \centering
    \includegraphics[width=0.47\textwidth]{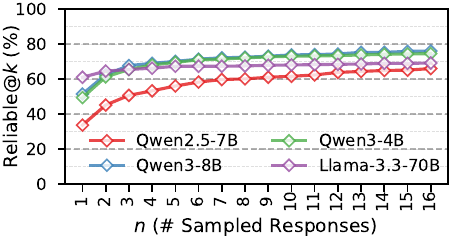}
    \vspace{-0.5em}
    \caption{\textbf{Scaling Landscape of Rejection Sampling}.}
    \label{fig:rejection-sampling}
    \vspace{-1em}
\end{figure}

\noindent $\bullet$ \textbf{Parallel (Rejection Sampling)}.
To scale test-time compute in parallel, we adopt sampling rather than greedy decoding, generating $n$ samples with a temperature of 1. We assume the availability of a response selector.
For example, in prompts related to \ourbench, format requirements can be extracted with the assistance of an LLM, while Python code can be used for rapid response evaluation.
With the response selector, the \reliableat{10} metric is scored positively if the 10 cousin prompts within one group all have at least one response that passes among the $n$ samples.
The results are shown in~\Cref{fig:rejection-sampling}. 
Rejection sampling with a suitable response selector effectively enhances the reliability of LLMs in instruction-following. 
The \reliableat{10} scores rise significantly as more samples are allowed, plateauing around $n=12$.
We further observe that reasoning models, such as \llm{Qwen3-8B}, benefit more from additional compute via parallel scaling, given their more exploratory nature~\citep{chen2025pass-at-k-training-lrm}. 
For instance, \llm{Qwen3-4B} and \llm{Qwen3-8B}, with only $n=3$ samples, already surpass \llm{LLaMA-3.3-70B} (the strongest open-source model in~\Cref{tab:main-results}). 
However, we note that extending this parallel scaling strategy to other domains may be challenging without suitable response selectors.

%









\section{Conclusion}

In this work, we present a systematic study on the nuance-oriented reliability of LLMs in instruction-following.
To characterize this property, we introduce the new metric \reliableatk, which can be instantiated with cousin prompts.
We design and implement a fully automated pipeline to generate such cousin prompts through data augmentation methods, combined with a code-assisted validity checker.
Leveraging this pipeline, we construct the \ourbench benchmark, which enables systematic evaluation.
Extensive experiments across 46 models using \ourbench uncover a critical vulnerability: current LLMs frequently fail to ensure reliability across cousin prompts with nuanced differences.
We further explore three directions for improving this reliability: prediction-based, training-based, and test-time scaling.
Overall, our findings highlight the importance of moving toward nuance-oriented reliability as a key dimension of dependable, trustworthy LLM behaviors.

\section{Acknowledgment}
This work was supported by Ant Group and the Center for High Performance Computing at Tsinghua University.


\section{Limitations}

We identify several points that are not fully solved in this work and are worth future investigation.

\noindent \textbf{Measurement Efficiency}.
Our practice of evaluating nuance-oriented reliability is straightforward, where we introduce cousin prompts to avoid the risk of hacking certain test cases.
This enables us to compute the \reliableatk metric to monitor the reliability dimension.
However, this also introduces a higher evaluation cost.
For example, a full evaluation of \ourbench requires 10× as many responses to be generated as \dataset{IFEval}.
Future work will focus on selecting or crafting those most differentiating test cases and cousin prompts, which enables a more cost-efficient evaluation.

\noindent \textbf{Missing Content-Level Evaluation}.
Following common practice in prior studies, we primarily evaluate whether LLMs adhere to constraint instructions.
Moreover, prompts often include one or multiple tasks, which are not explicitly assessed by either \dataset{IFEval} or \ourbench.
This design choice is acceptable, as satisfying constraints is a necessary prerequisite for correctly following the entire prompt.
Future work could extend this by jointly measuring content-level response quality along with constraint adherence.

\noindent \textbf{Focus on \dataset{IFEval}}.
In this work, we primarily adopt \dataset{IFEval} as our testbed.
It is widely recognized in the community as a standard benchmark for instruction-following evaluation and has approached performance saturation with the latest models.
Nevertheless, our reliability evaluation methodology is not limited to this single benchmark.
Future research may extend our approach to other existing benchmarks, or even beyond the instruction-following domain.
We envision that evaluating nuance-oriented reliability will become a default practice, complementing traditional capacity-oriented evaluation.

\noindent \textbf{Potential Bias in Automated Validity Check}.
Although the code-assisted validity checker demonstrates high recall and precision, its automation depends on LLM-based judgment, which may introduce subtle biases or misclassifications. 
We carefully and systematically validate the checker’s performance in~\Cref{subsec:checker-design} and additionally dedicate substantial, though undocumented, manual effort to continuously monitor and ensure the quality of the checking process.
Moreover, the strong performance of advanced models---such as \llm{GPT-5}, which achieves a \reliableat{10} of 78.4---suggests that most test cases are indeed valid and feasible; otherwise, invalid cases would cause even the strongest LLMs to fail consistently.

\noindent \textbf{Limited Coverage of Improvement Recipes}.
While numerous efforts have been proposed to enhance instruction-following capabilities, reproducing all such methods to assess their combined impact on nuance-oriented reliability lies beyond the scope of this work. 
We focus on several representative improvement strategies, including prediction-based, training-based, and test-time scaling approaches. 
Future research could undertake a more comprehensive investigation to identify the most effective techniques for improving reliability across diverse enhancement paradigms.

\noindent \textbf{Cross-Lingual Reliability}.
Our study primarily evaluates instructions expressed in English, leaving the applicability of \textit{reliable@k} to under-resourced languages largely unexplored. 
In principle, the metric can be adapted through practical adjustments, such as incorporating a translation module in the data augmentation pipeline and extending evaluation functions to additional languages. 
Assessing reliability across languages expands the scope of model evaluation and may reveal critical insights into the fairness and generalizability of LLM performance for users of diverse linguistic backgrounds. 
We envision that a comprehensive investigation of cross-lingual reliability will yield further insights into model generalization.

\section{Ethical Considerations}

Our research adheres to the ethical guidelines established by the Association for Computational Linguistics (ACL)\footnote{\url{https://aclrollingreview.org/responsibleNLPresearch/}}. 
We have made every effort to conduct this work with the utmost respect for ethical principles and research integrity.

This research aims to advance the understanding of the reliability of large language models in instruction-following. 
All experiments were performed on publicly available models and datasets, ensuring compliance with relevant terms of service and data usage policies, \eg, \dataset{IFEval}~\citep{zhou2023instruction-ifeval-verification-instruction-following-evaluation-25-types}.
No proprietary or confidential information was accessed or reverse-engineered during this study.
Our analyses do not involve human subjects, sensitive personal data, or the generation of harmful content.
Our use of AI assistants is limited to writing polishing and grammar checking.
We release the complete codebase and datasets (\eg, \ourbench) to ensure full reproducibility of our results. 

Finally, we remind the readers that any techniques introduced in this paper should be applied ethically and within appropriate research contexts.

\bibliography{refs,refs-if}

\clearpage
\appendix
\crefalias{section}{appendix}
\crefalias{subsection}{appendix}
\section{Related Works}\label{appx:related-works}

\subsection{Benchmarking Instruction-Following}
\label{appx:if-benchmarks}

\begin{table*}
    \caption{\textbf{Key Designs of the Existing Instruction-Following Benchmarks}.}
    \label{tab:if-benchmark-taxonomies}
    \centering
    \large
    \resizebox{\textwidth}{!}{
    \begin{tabular}{lccccc}
    \toprule
    \multirow{2}{*}{\textbf{Benchmark}} & \multirow{2}{*}{\textbf{\# Cases}} &  \multicolumn{3}{c}{\textbf{Comprehensiveness-Oriented}} & \multirow{2}{*}{\textbf{Nuance-Oriented}}  \\
     &  & New Tasks & New Constraints & Complicated Scen. & \\
    \midrule
     \dataset{Collie-v1}~\citep{yao2024collie-constrained-text-generation} & 2,080 & \LEFTcircle & \CIRCLE & \Circle & \Circle \\
      \dataset{CELLO}~\citep{he2024can-large-language-models-understand-real-world-complex-instructions-aaai-cello} & 523 & \Circle & \LEFTcircle & \CIRCLE & \Circle \\
      \dataset{NPB}~\citep{sun2023evaluating-controlled-generation-tasks} & -- & \CIRCLE & \CIRCLE & \Circle & \Circle \\
      \dataset{FollowBench}~\citep{jiang2024followbench-multi-constraints-composition-and-considering-decoding-methods} & 820 & \LEFTcircle & \CIRCLE & \Circle & \Circle\\
      \dataset{IFEval}~\citep{zhou2023instruction-ifeval-verification-instruction-following-evaluation-25-types} &  541 & \Circle & \CIRCLE & \Circle & \Circle\\
     \dataset{CoDI-Eval}~\citep{chen2024benchmarking-CoDI-Eval-aaai-diversified-instructions} & 9,060 & \Circle & \CIRCLE & \Circle & \LEFTcircle\\
     \dataset{InFoBench}~\citep{qin2024infobench-evaluate-with-constraint-embedded-and-eval-via-binary-questions} & 500 & \Circle & \CIRCLE & \LEFTcircle & \Circle \\
     \dataset{FOFO}~\citep{xia2024fofo-a-benchmark-to-evaluate-llms-format-following-capability-acl-long-papers} & 494 & \CIRCLE & \CIRCLE & \Circle & \Circle \\
      \dataset{AlpacaEval-LI}~\citep{yuan2024following-length-constraints-in-instructions} & 802  & \Circle & \CIRCLE & \LEFTcircle & \Circle \\
      \dataset{MT-Bench-LI}~\citep{yuan2024following-length-constraints-in-instructions} & 240  & \Circle & \CIRCLE & \LEFTcircle & \Circle \\
      \dataset{SIFo}~\citep{chen2024sifo-sequential-instruction-following} & 800 & \LEFTcircle & \Circle & \Circle & \CIRCLE \\
      \dataset{RuleBench}~\citep{sun2024beyond-inferential-rule-following-rulebench} & --   & \CIRCLE & \Circle & \CIRCLE & \Circle\\
      \dataset{ComplexBench}~\citep{wen2024benchmarking-complexbench-multiple-constraints-composition} & 1,150 & \Circle & \LEFTcircle & \CIRCLE & \Circle\\
      \dataset{CFBench}~\citep{zhang2024cfbench-comprehensive-constraints-following-benchmark-for-llms} & 1,000  & \LEFTcircle & \CIRCLE & \CIRCLE  & \Circle\\
      \dataset{SysBench}~\citep{qin2024sysbench-system-message-following-benchmark-with-constraints}& 500  & \Circle &\Circle  & \CIRCLE & \Circle\\
      \dataset{MMMT-IF}~\citep{epstein2024mmmt-if-multimodal-multi-turn-instruction-following}& 71  & \CIRCLE & \LEFTcircle & \CIRCLE &\Circle \\
      \dataset{RealInstruct}~\citep{ferraz2024llm-realinstruct-real-world-instruction-following-emnlp} & 302 & \Circle & \CIRCLE & \Circle & \Circle\\
      \dataset{Multi-IF}~\citep{he2024multi-if-multi-turn-and-multilingual-instructions-following-meta} & 4,501 & \Circle &\Circle & \CIRCLE  &\Circle \\
      \dataset{LIFBench}~\citep{wu-etal-2025-lifbench-instruction-following-in-long-context} & 2,766  & \Circle & \LEFTcircle & \CIRCLE & \Circle \\
      \dataset{VFF}~\citep{wang2025verifiable-vff-naacl-format-control} & 21k & \Circle & \CIRCLE & \Circle & \Circle\\
      \dataset{StructFlowBench}~\citep{li2025structflowbench-structured-flow-benchmark-for-multi-turn-instruction-following} & 155 & \Circle & \Circle & \CIRCLE & \Circle\\
      \dataset{PBIF}~\citep{zeng2025order-position-bias-in-multi-constraint-instruction-following} & 24k & \Circle & \Circle & \Circle & \CIRCLE \\
      \dataset{CodeIF}~\citep{yan2025codeif} & 1,200 & \CIRCLE & \CIRCLE & \LEFTcircle & \Circle\\
      \dataset{WildIFEval}~\citep{lior2025wildifeval} & 11,813 & \Circle  & \CIRCLE & \LEFTcircle & \Circle\\
      \dataset{XIFBench}~\citep{li2025xifbench-multilingual-instruction-following} & 465  & \Circle & \LEFTcircle & \CIRCLE & \Circle\\

      \dataset{CodeIF-Bench}~\citep{wang2025codeif-bench-interactive-code-generation} & -- & \Circle & \CIRCLE & \CIRCLE & \Circle\\
      \dataset{MulDimIF}~\citet{ye2025multi-dimensional-constraint-framework-fudan} &1,200& \Circle& \CIRCLE & \LEFTcircle & \Circle\\
      \dataset{MathIF}~\citep{fu2025scalingreasoninglosingcontrol} & 420 & \CIRCLE & \LEFTcircle & \CIRCLE & \Circle\\
      \dataset{AGENTIF}~\citep{qi2025agentifbenchmarkinginstructionfollowing}& 707 & \CIRCLE & \CIRCLE & \CIRCLE & \Circle \\
      \dataset{LIFEBench}~\citep{zhang2025lifebench-evaluating-length-instruction} & 10,800 & \LEFTcircle & \CIRCLE & \Circle & \Circle\\
      \dataset{MaXIFE}~\citep{liu2025maxifemultilingualcrosslingualinstruction} & 18,285 & \Circle & \Circle  & \CIRCLE & \Circle\\
      \dataset{EifBench}~\citep{zou2025eifbench} & 1,000 & \LEFTcircle & \LEFTcircle & \CIRCLE &\Circle \\

      \dataset{Ordered CommonGen}~\citep{song2025ifircomprehensivebenchmarkevaluating} & 27,648 & \CIRCLE & \CIRCLE & \LEFTcircle & \Circle \\
      \dataset{IFBench}~\citep{pyatkin2025generalizing-if-bench} & 300  & \Circle & \CIRCLE & \Circle & \Circle \\
      \dataset{IFScale}~\citep{jaroslawicz2025instructionsllmsfollowonce} & -- & \Circle & \CIRCLE & \LEFTcircle & \LEFTcircle\\
      \dataset{IFEvalCode}~\citep{yang2025ifevalcodecontrolledcodegeneration} & 1,600 & \CIRCLE & \LEFTcircle & \LEFTcircle & \Circle \\
      
    \bottomrule
    \end{tabular}}
\end{table*}

We provide a chronological survey in~\Cref{tab:if-benchmark-taxonomies}.

\dataset{Collie-v1}~\citep{yao2024collie-constrained-text-generation}.
This work proposes a grammar-based framework for constructing controlled text generation, allowing automatic conversion from formal constraint configurations to natural language instructions.
The final compiled benchmark includes 2,080 instances comprising 13 constraint structures.

%
\dataset{CELLO}~\citep{he2024can-large-language-models-understand-real-world-complex-instructions-aaai-cello}.
Accounting for the complexity of in-the-wild user instructions, this work focuses on long and complex instructions.  
The proposed \dataset{CELLO} benchmark contains 523 instructions that cover four types of response constraints: count limit, answer format, input dependency, and phrase callback.

\dataset{NPB}~\citep{sun2023evaluating-controlled-generation-tasks}. 
This work evaluates the controllability of LLMs during generation tasks. It includes numerical planning constraints, requesting responses with varying lengths.

\dataset{FollowBench}~\citep{jiang2024followbench-multi-constraints-composition-and-considering-decoding-methods}.
This work assesses the constraint-following abilities of LLMs, considering five types of constraints: content, situation, style, format, and example.
It involves incrementally adding constraints when curating instructions, yielding a total of 820 instructions with different difficulty levels.

\dataset{IFEval}~\citep{zhou2023instruction-ifeval-verification-instruction-following-evaluation-25-types}.
This work pioneers in proposing a broad taxonomy of code-verifiable constraints.
The resulting benchmark consists of 541 instructions, covering 25 types of constraints.

\dataset{CoDI-Eval}~\citep{chen2024benchmarking-CoDI-Eval-aaai-diversified-instructions}.
This work introduces expansion and rewriting steps to enhance instruction diversity. The resulting \dataset{CoDI-Eval} benchmark consists of 9,060 instructions and covers a range of constraints, including sentiment, topic, keyword, length, and toxicity avoidance.

\dataset{InFoBench}~\citep{qin2024infobench-evaluate-with-constraint-embedded-and-eval-via-binary-questions}. 
This work proposes a novel scoring method: preparing a checklist for each instruction and employing LLM-as-a-Judge to decide the acceptance.
Meanwhile, it introduces the \dataset{InFoBench} benchmark, which comprises 500 instructions and involves constraints like content constraint, linguistic constraint, style rules, format specifications, and number limits.

\dataset{FOFO}~\citep{xia2024fofo-a-benchmark-to-evaluate-llms-format-following-capability-acl-long-papers}. 
This work specifically studies the format-following abilities of LLMs. It proposes the \dataset{FoFo} benchmark, which contains 494 instructions. Each test case requires the responses to adhere to domain-specific formats, \eg, medical reports and Latex. LLM-as-a-Judge is employed for automated adherence evaluation.

\dataset{AlpacaEval-LI \& MT-Bench-LI}~\citep{yuan2024following-length-constraints-in-instructions}. 
This work investigates LLMs' ability to follow length constraints in instructions. 
To facilitate this evaluation, the authors extend \dataset{AlpacaEval}~\citep{dubois2024length-controlled-alpacaeval} and \dataset{MT-Bench}~\citep{zheng2024llm-judge-fastchat} by constructing the corresponding length-instructed variants

\dataset{SIFo}~\citep{chen2024sifo-sequential-instruction-following}.
This work focuses on a specific aspect of instruction-following evaluation---ordering of multiple instructions (\ie, constraints in this work).
The \dataset{SIFo} benchmark comprises 800 test cases, with LLMs explicitly instructed to address multiple constraints in a prescribed sequential order.
This is implemented as a single-turn interaction.

\dataset{RuleBench}~\citep{sun2024beyond-inferential-rule-following-rulebench}.
This work studies inferential rule following, where additional \textit{if–then} rules are incorporated into the instructions.
This setting differs from the affirmatively requested constraints in other works.

\dataset{ComplexBench}~\citep{wen2024benchmarking-complexbench-multiple-constraints-composition}.
This work studies LLMs' abilities to follow complex instructions involving multiple constraint compositions.
The \dataset{ComplexBench} consists of 1,150 instructions, each associated with a constraint composite.
The composition types considered include \textit{And}, \textit{Chain}, \textit{Selection}, and \textit{Nested}.

\dataset{CFBench}~\citep{zhang2024cfbench-comprehensive-constraints-following-benchmark-for-llms}.
This work emphasizes the comprehensiveness of constraint types for evaluating instruction following.
It proposes a constraint system with 25 subcategories and meticulously curates 1K Chinese instructions.

\dataset{SysBench}~\citep{qin2024sysbench-system-message-following-benchmark-with-constraints}.
This work echoes the real-world user practices involving system prompts.
They evaluate the instruction-following ability of LLMs, involving 500 system prompts requesting six types of constraints.

\dataset{MMMT-IF}~\citep{epstein2024mmmt-if-multimodal-multi-turn-instruction-following}.
This work extends the instruction-following evaluation to multi-modal and multi-turn settings.
The \dataset{MMMT-IF} benchmark contains 71 test cases, with image inputs and multiple constraints scattered across chat.

\dataset{RealInstruct}~\citep{ferraz2024llm-realinstruct-real-world-instruction-following-emnlp}.
This paper introduces real-world user queries as a source for instruction-following test cases.
The authors meticulously curate instruction candidates from ShareGPT and then decompose them into tasks, contexts, and constraints.
The final test set includes 302 instances, each of which typically contains multiple constraints.

\dataset{Multi-IF}~\citep{he2024multi-if-multi-turn-and-multilingual-instructions-following-meta}.
This work extends instruction-following evaluation to multi-turn and multilingual settings.
The \dataset{Multi-IF} benchmark is composed of 4,501 multilingual conversations, constructed by augmenting \dataset{IFEval} with multi-turn interactions and translations.

\dataset{LIFBench}~\citep{wu-etal-2025-lifbench-instruction-following-in-long-context}.
This work studies instruction-following abilities of LLMs in long-context tasks, \eg, processing multiple documents.
The benchmark consists of 2,766 instructions, involving long-context tasks and 6 types of instruction-following capabilities.

\dataset{VFF}~\citep{wang2025verifiable-vff-naacl-format-control}.
This work exclusively explores the format-constraint following abilities of LLMs.
It constructs 21K test cases with varying levels of difficulty by combining one to three verifiable constraints per instruction. 
The constraint types include limited word count, limited content, limited punctuation, limited structure, limited grammar, and specific number formatting.

\dataset{StructFlowBench}~\citep{li2025structflowbench-structured-flow-benchmark-for-multi-turn-instruction-following}.
This work explores the instruction-following performance of LLMs in multi-turn scenarios, emphasizing the structural dependencies that arise across turns.
The benchmark incorporates eight types of constraints and six categories of structural flows: \textit{follow-up}, \textit{refinement}, \textit{recall}, \textit{expansion}, \textit{summary}, and \textit{unrelatedness}. In total, it includes 155 dialogues, evaluated via LLM-as-a-Judge.

\dataset{PBIF}~\citet{zeng2025order-position-bias-in-multi-constraint-instruction-following}.
This work investigates how the ordering of multiple constraints influences the instruction-following performance of LLMs. 
It identifies an intriguing phenomenon: LLMs often perform better when the constraints are presented in a \textit{hard-to-easy} order.

\dataset{CodeIF}~\citep{yan2025codeif}.
This work introduces a benchmark for evaluating LLMs' instruction-following in code generation, covering tasks like synthesis, debugging, refactoring, and explanation.  
It comprises 1,200 test cases across 8 categories and 50 sub-instructions, spanning multiple programming languages.

\dataset{WildIFEval}~\citep{lior2025wildifeval}.
This work curates a set of 12K complex instructions collected from real-world user queries on Chatbot Arena~\citep{chiang2024chatbot-arena-an-open-platform-for-evaluating-llms-by-human-preference-icml}.
Each instruction is then decomposed into 8 categories of constraints, and LLM-as-a-Judge is employed for automated evaluation of instruction-following.

\dataset{XIFBench}~\citep{li2025xifbench-multilingual-instruction-following}.
This work primarily explores the instruction-following abilities of LLMs across different language settings, from high-resource (English and Chinese) to low-resource (Hindi).
It starts with five categories of constraints: content, style, situation, format, and numerical.
By composing between one and five constraints into each seed instruction, the authors construct a total of 465 hard instructions. 
These are further extended into multiple languages through translation, resulting in a multilingual benchmark.

\dataset{CodeIF-Bench}~\citep{wang2025codeif-bench-interactive-code-generation}.
This work focuses on multi-turn interactive code generation.  
It covers 9 types of verifiable instructions aligned with real-world software development requirements for test-case validation, and supports both Static and Dynamic Conversation settings.  


\dataset{MulDimIF}~\citet{ye2025multi-dimensional-constraint-framework-fudan}.
This work proposes a multi-dimensional constraint framework, involving three types of constraint formatting patterns (\textit{Example}, \textit{Listing}, and \textit{Incorporation}), four categories of constraints, and four difficulty levels.
In total, this benchmark consists of 1,200 code-verifiable instructions.

\dataset{MathIF}~\citep{fu2025scalingreasoninglosingcontrol} is a dedicated benchmark for evaluating instruction-following in mathematical reasoning tasks.
It contains 420 evaluation samples derived from 15 Python-verifiable constraints across 4 categories, applied to math problems of varying difficulty.

\dataset{AGENTIF}~\citep{qi2025agentifbenchmarkinginstructionfollowing} evaluates LLMs’ instruction-following abilities in agentic scenarios with long, complex, and constraint-heavy instructions.  
It comprises 707 human-annotated instructions drawn from 50 real-world agentic tasks.

\dataset{LIFEBench}~\citep{zhang2025lifebench-evaluating-length-instruction}.
This work is conducted with a focus on length constraints, ranging from 16 to 8192 words.
It contains 10,800 instructions, in English or Chinese, requesting responses with length satisfying three types of relations (\textit{At Most}, \textit{At Least}, and \textit{Equal To}).

\dataset{MaXIFE}~\citep{liu2025maxifemultilingualcrosslingualinstruction}.
This work focuses on multilingual and cross-lingual scenarios.  
It consists of parallel data in 23 languages, with each record combining a Basic Question and 1–3 Instructions, totaling 18,285 entries.

\dataset{EifBench}~\citep{zou2025eifbench} introduces a new evaluation paradigm---multi-instruction, multi-constraint---to assess the instruction-following capabilities of LLMs, reflecting the multifaceted requirements often encountered in real-world scenarios.
The benchmark contains 1,000 instances, each featuring a complex scenario description accompanied by a set of $N$ tasks, each governed by $M$ constraints.
For each instance, the model is required to respond to all $N \times M$ instructions.

\dataset{Ordered CommonGen}~\citep{song2025ifircomprehensivebenchmarkevaluating} introduces a benchmark that evaluates LLMs’ ability to generate concepts in a specified order.
It contains 27,648 instances from 192 concept sets and 6 instruction templates, using ordered coverage to measure adherence to the specified sequence.

\dataset{IFBench}~\citep{pyatkin2025generalizing-if-bench}.
This work highlights the limited generalization of LLMs' instruction-following abilities across constraint types, thus falling into comprehensiveness-oriented reliability.
They curate a new set of 300 instructions, involving 58 new constraint types beyond the 25 types in \dataset{IFEval}.

\dataset{IFScale}~\citep{jaroslawicz2025instructionsllmsfollowonce}.
This work introduces a benchmark for evaluating instruction-following performance as instruction density increases in professional business report generation.  
Each task contains 10–500 keyword-inclusion instructions drawn from a curated 500-term business vocabulary, with density increasing in steps of 10.

\dataset{IFEvalCode}~\citep{yang2025ifevalcodecontrolledcodegeneration} introduces a multilingual benchmark for controlled code generation, evaluating LLMs' ability to follow instructions beyond correctness in coding scenarios. 
It includes 1,600 samples in 8 languages with paired Chinese–English queries, assessing both correctness and instruction adherence.

Among these, several works are relevant to the techniques and the research problem studied in this work.
We distinguish ours from several highly relevant ones:
\begin{packeditemize}
\item \dataset{CoDI-Eval}~\citep{chen2024benchmarking-CoDI-Eval-aaai-diversified-instructions}:
This benchmark mainly focuses on the controllability of LLMs in generation tasks.
This work advocates an instruction diversification process to synthesize diverse forms of constraint expression (\ie, the rewriting step).
They aim to achieve finer-grained coverage of instructions rather than to study the reliability issue.
What's more, they fail to account for inter-prompt relationships.
\item \dataset{SIFo}~\citep{chen2024sifo-sequential-instruction-following} \& \dataset{PBIF}~\citep{zeng2025order-position-bias-in-multi-constraint-instruction-following}:
These two works investigate the position bias problem of LLMs in the context of instruction-following.
They examine how LLMs' compliance changes when the order of constraints within prompts is altered.
This phenomenon can be viewed as a special case of our rephrasing-based augmentation approach.
\item
\dataset{LIFEBench}~\citep{zhang2025lifebench-evaluating-length-instruction}:
This work specifically studies length constraints.
Their study is analogous to our pilot study in~\Cref{subsec:problem-statement}.
However, they focus on how long-context LLMs follow length constraints rather than on the general reliability. 
\item \dataset{IFScale}~\citep{jaroslawicz2025instructionsllmsfollowonce}:
This work studies how many instructions LLMs can simultaneously follow.
For the research question, the study also relates to reliability, but it stress-tests LLMs along the complexity-oriented dimension.
\end{packeditemize}

\subsection{Evaluating LLMs' Reliability}
\label{appx:related-work-reliability}

There exist works revealing LLMs' sensitivity to prompt wordings~\citep{sclar2024quantifying-how-i-learned-to-start-worrying-about-prompt-formatting-classification-tasks}.
\citet{cao2024worst-prompt-performance-of-llms} specifically study the worst performance across semantically equivalent case-level queries.
\citet{mizrahi2024state-multi-prompt-llm-evaluation} advocate for multi-prompt LLM evaluation, which shares the same methodology as our curation of cousin prompts.
Based on these works, we move beyond studying simply prompt sensitivity issues (the rephrasing augmentation may reflect) and towards more general nuance-oriented reliability.

A critical distinction must be drawn between the \reliableatk metric and sampling responses for one prompt independently $k$ times.
\citet{yao2025taubench-tool-calling-agent} introduce a new metric, \texttt{pass\^{}k}, for evaluating agentic scenarios. 
While \texttt{pass\^{}k} measures reliability across $k$ independent trials using the \textit{same} prompt (often with temperature $>0$), our \reliableatk metric evaluates consistency across \textit{distinct} cousin prompts with nuanced variations. 
To quantitatively compare these two dimensions, we conducted an experiment sampling each \dataset{IFEval} prompt 10 times at temperature 1 to compute \texttt{pass\^{}10}, contrasting it with our \reliableatk on \ourbench.

\begin{table}[!ht]
    \centering
    \caption{Comparison between standard Accuracy, \reliableat{10}, and \texttt{pass\^{}10}.}
    \label{tab:reliable_vs_pass}
    \resizebox{0.48\textwidth}{!}{
    \begin{tabular}{lccc}
        \toprule
        \textbf{Model Name} & \textbf{Accuracy} & \textbf{\reliableat{10}} & \textbf{\texttt{pass\^{}10}} \\
        & (IFEval) & (IFEval++) & (Repeated Sampling) \\
        \midrule
        Qwen2.5-7B-Instruct & 73.0 & 34.8 & 53.0 \\
        Qwen3-4B & 85.2 & 52.5 & 67.0 \\
        Qwen3-8B & 87.6 & 58.8 & 71.0 \\
        Llama-3.3-70B-Instruct & 92.1 & 71.0 & 85.6 \\
        \bottomrule
    \end{tabular}
    }
    \vspace{-1em}
\end{table}

As shown in~\Cref{tab:reliable_vs_pass}, \reliableat{10} and \texttt{pass\^{}10} degrade at different rates compared to standard accuracy.
For instance, even for strong models like \llm{LLaMA-3.3-70B}, the gap between single-prompt accuracy (92.1) and cousin-prompt reliability (71.0) is significant, and distinct from the gap caused by sampling variance (85.6).
This highlights that they capture different aspects of LLMs in instruction-following: \texttt{pass\^{}k} reflects stability against random seed noise, whereas \reliableatk reflects robustness against semantic nuances.
Therefore, the two evaluation methods are orthogonal and can be jointly used to characterize the lower bound of an LLM's reliability more rigorously.
Furthermore, as we discuss in~\Cref{subsec:test-time-scaling}, the variability exposed by multiple samples (whether from repeated sampling or cousin prompts) can be exploited via rejection sampling to improve final performance.

\section{Data Adaption}
\label{appx:data-adaption}

\subsection{Constraint Types}
\label{appx:constraint-type}
We adopt the 541 test cases from~\dataset{IFEval}.
All of them are composed of tasks with several format constraints.
The format constraints are categorized into 25 types, which we detail in~\Cref{tab:ifeval-constraints-type-part1}.
These \textbf{simple} constraints lay the foundation for LLMs' solving various more complex requirements.
As most LLMs,  as reported by the community and verified by our experiments (the \textit{ACC} column in~\Cref{tab:main-results}), have boosted performance in the original \dataset{IFEval}, it is a necessary next step to understand how solid and reliable their excellent instruction-following abilities are.
This motivates the nuance-oriented reliability examination.

\subsection{Principles for High-Quality Data}
\label{appx:checker-principles}

Recall that each test case mainly relates to three key components:
the prompt which instructs LLMs, the evaluation function for checking the instruction-following behaviors (specifically, Python code and response formats in the context of \dataset{IFEval} and \ourbench), and the evaluation configurations which specify how the evaluation functions should examine the instruction-following behaviors.

We document four principles for high-quality test cases, which are also enforced in our data augmentation in~\Cref{subsec:data-augmentation} and implemented in our automated validity checker (see~\Cref{prompt:sanity-checker}).
\begin{packeditemize}
\item \textbf{Clear Instruction Intents}:
Each test case must contain an unambiguous and self-contained prompt that clearly specifies the task to be performed and the constraints to be satisfied. 
The prompt should minimize interpretation uncertainty and avoid overlapping objectives (\eg, mixing summarization with classification). 
This ensures that the model's output can be attributed to understanding and following the instruction rather than guessing the task’s intent.

\item \textbf{Verifiability through Evaluation Implementations}:
A high-quality test case must be verifiable by automated evaluation functions.
The expected outputs or behavioral criteria should be formulated so that they can be programmatically validated without requiring human judgment.
Since certain evaluation function implementations in \dataset{IFEval} possess inherent assumptions and inductive biases, the specified constraint requirements must remain within the boundaries of these evaluation functions.
Meanwhile, the prompts must articulate the constraint requirements in a way that the downstream evaluation functions can verify them.

\item \textbf{Alignment with Evaluation Configurations}:
The evaluation configuration (\eg, the \textit{relation} and \textit{num\_words} for the length constraints) must be compatible with both the prompt and the evaluation function. 
Misalignments between prompt requirements and evaluation logic can produce false positives or negatives, undermining the reliability of the benchmark. 
Thus, each test case should be explicitly validated against the intended constraint requirements (\ie, the evaluation configurations) to ensure consistency.

\item \textbf{Attainability}:
A test case should be solvable by a capable model based solely on the given prompt.
Consider a scenario where two length constraints coexist within a single test case:
one requiring more than ten sentences and another requiring fewer than eight words.
Such constraints are inherently incompatible, rendering the test case infeasible and therefore subject to removal.
We observe that \dataset{IFEval} includes several test cases exhibiting these attainability issues.
Moreover, since our data augmentation strategies---namely, adding distractors and altering constraints---may occasionally combine multiple constraints, we design and implement our validity checker to carefully filter out any combinations that result in unattainable or contradictory requirements.

\end{packeditemize}

\subsection{Revising Infeasible Test Cases}
\label{appx:revise-test-cases}

Manual review and community feedback reveal that a small number of test cases in \dataset{IFEval} are flawed.
We make necessary yet minimal revisions to ensure test case integrity.
The overall information and examples are listed in~\Cref{tab:revision-examples}.
We identify that common issues in \dataset{IFEval} can be categorized into the following four classes.

\noindent \textbf{Flawed Implementation of Evaluation Functions}.
A representative issue arises from mismatches between prompts and the underlying automatic checkers. For example, one test case requires the ``\#'' character to appear at least four times. 
However, the original evaluation function only supports letters a-z, making it impossible to evaluate such symbols. As a result, this prompt is invalid for the evaluation function, despite being well-formed.
We updated the evaluation function to accept any character—including symbols and digits---so that constraints targeting non-alphabetic characters can be correctly evaluated, ensuring the evaluation function aligns with the intended instructions.

\noindent \textbf{Flawed Documentation of Evaluation Configurations}.
In some instances, the natural language prompt directly contradicts the evaluation arguments. 
For example, one prompt requires ``the letter o to appear at least 6 times'', but the evaluation argument specifies ``less than 6 times. 
Such inconsistencies lead to an unavoidable conflict between what the prompt demanded and how correctness is assessed. 
We corrected these cases by rewriting the prompt to match the evaluation arguments.

\noindent \textbf{Typos in Natural Language Prompts}.
Some test cases contain minor typographical errors that make them misleading. 
For instance, one prompt instructs to separate two tables using ``exactly 6 asterisk symbols: *******'', where the example actually has 7 asterisks, contradicting the requirement. 
This was revised to six asterisks ``******'' to match the stated constraint, thereby eliminating ambiguity introduced by typographical mistakes.

\noindent \textbf{Unachievable Multi-Constraint Test Cases}.
Some prompts impose multiple constraints that are logically incompatible. 
A notable example requires the model first to repeat a sentence verbatim (including commas) and then simultaneously adhere to the instruction ``avoid using commas''. 
Since the two conditions could not be satisfied simultaneously, such prompts were fundamentally unachievable. 
We resolved this by removing the contradictory requirement, keeping the prompt feasible.


\section{Model Details}
\label{appx:model-details}
In this work, we include a total of 36 large language models to establish a representative benchmark. Our selection aims to comprehensively capture the development trends of LLMs in terms of reliability. 
To this end, we include models that are widely recognized for their performance or influence within the research and developer communities, covering a diverse range of providers, architectures, and scales.
For those proprietary models, we access their LLM services via API, specifically the OpenAI-compatible interface.
For the open-source ones, we locally deploy them using vLLM~\citep{kwon2023vllm} for inference.
We use the instruction-tuned or chat version and adopt official chat templates.
See~\Cref{tab:model-details} for the detailed information.

\begin{figure*}[t]
    \centering
    \includegraphics[width=\linewidth]{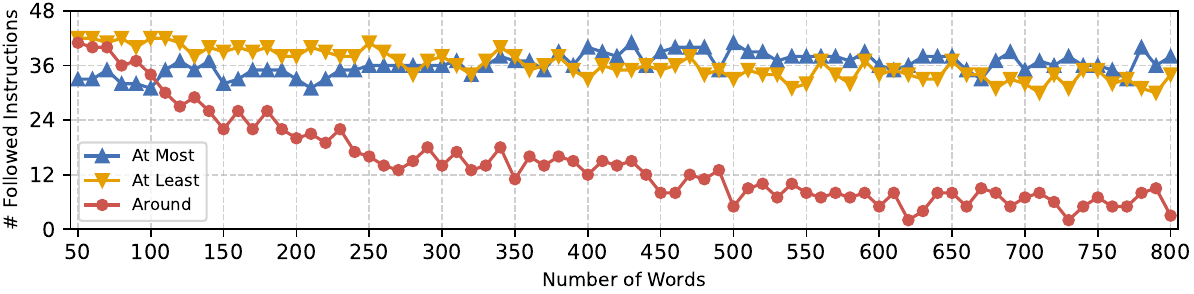}
    \vspace{-2em}
    \caption{\textbf{Results of Requesting Varying Number of Response Words}. Experiments with \llm{Qwen3-32B}.}
    \label{fig:poc-results-qwen3-32b}
    \vspace{-1em}
\end{figure*}

\section{More Experimental Results}
\label{appx:more-experiments}

\subsection{Impact of the System Prompt}
\label{appx:system-prompt-impact}

Certain LLMs are published with default system prompts.
For instance, models in the \llm{Qwen2.5} series include default prompts in their chat templates. 
In contrast, more recent reasoning models, such as \llm{DeepSeek-R1} and \llm{Qwen3}, do not specify system prompts; their providers even recommend using an empty prompt.
To elicit the default (and arguably optimal) behaviors while avoiding evaluation bias, our main experiments adopt the system prompt settings recommended by model providers.

Meanwhile, we notice that it is a common practice for service providers or end users to employ system prompts tailored for personalized LLM services~\citep{ge2024scaling-persona-hub,qin2024sysbench-system-message-following-benchmark-with-constraints}.
We are also curious about the impact of system prompt setting on the models' reliability in instruction-following.
We include six settings as listed in~\Cref{tab:system-prompt-settings}:
the default, three persons sampled from~\citet{ge2024scaling-persona-hub}, python coder, and encouraging better instruction-following.
Except for the system prompt setting, other decoding strategies remain consistent.

\begin{figure*}
    \centering
    \includegraphics[width=1\linewidth]{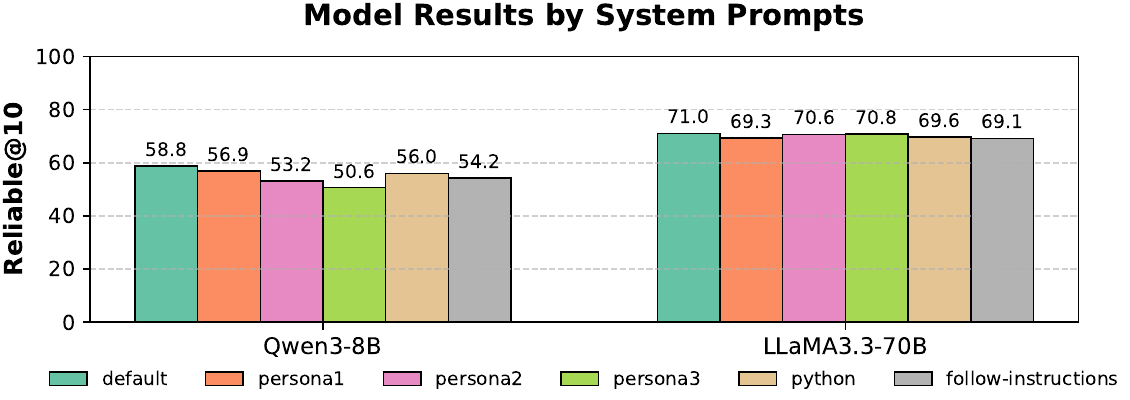}
    \vspace{-1em}
    \caption{\textbf{Impact of System Prompts on Nuance-Oriented Reliability}.}
    \label{fig:system-prompt-impact}
\end{figure*}

The results, illustrated in~\Cref{fig:system-prompt-impact}, show that system prompts have an implicit but notable impact on performance in \ourbench. 
Empirically, the default system prompt tends to yield the best instruction-following performance, confirming the representativeness of our choice in the main experiments. 
In contrast, personalization may harm reliability: certain persona prompts reduce \reliableat{10} by up to 8.2\%. 
Moreover, explicitly encouraging better instruction-following (the \textit{follow-instructions} prompt) does not improve performance. 
Importantly, the same prompt may affect models differently---the \textit{persona3} prompt significantly degrades the reliability of \llm{Qwen3-8B}, but has little or no impact on \llm{LLaMA-3.3-70B}.

This initial empirical measurement substantiates the validity of our choice of using the default system prompt.
However, the mechanisms by which the setting of system prompts implicitly influence downstream tasks such as \ourbench remain poorly understood, highlighting an important direction for future research.

\subsection{Scalability of the \reliableatk Metric}
\label{appx:relatk-scalability}

The \reliableatk metric is inherently configurable, depending on the choice of $k$ and the underlying cousin prompts. 
By increasing $k$, we can construct more challenging test cases, as the LLM must handle a larger set of cousin prompts simultaneously to achieve a score. 

\begin{figure}[t]
    \centering
    \includegraphics[width=0.47\textwidth]{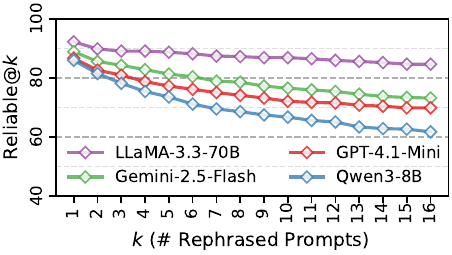}
    \vspace{-0.5em}
    \caption{\textbf{The \reliableatk Metric is Highly Scalable}.}
    \label{fig:scalability-of-reliable-at-k}
    \vspace{-1em}
\end{figure}

As illustrated in~\Cref{fig:scalability-of-reliable-at-k}, simple augmentation via rephrasing (strictly following the automated validity checker described in~\Cref{subsec:checker-design}) can steadily raise the difficulty up to approximately $k=16$.
Two key observations emerge: 
1) Increasing $k$ can better differentiate the reliability of different LLMs. 
For instance, \llm{Qwen3-8B} and \llm{GPT-4.1-mini} exhibit similar reliability when $k$ is small, but as $k$ grows, \llm{GPT-4.1-mini} demonstrates higher reliability than \llm{Qwen3-8B}. 
2) This highlights a promising pathway for scalable oversight. 
Employing sophisticated augmentation methods, such as distractors, can further enhance the scalability of the evaluation.

An additional implicit advantage of this approach, particularly for nuance-oriented evaluation, is data efficiency: test cases can be augmented directly within existing benchmarks, following the recipe established in this work, without requiring large-scale new data collection.

\begin{table*}[t]
    \centering
    \caption{\textbf{The Category-Wise Performance Decrease of LLMs.}}
    \label{tab:category_wise_performance}
    \resizebox{0.98\textwidth}{!}{
    \begin{tabular}{lcccccccc}
        \toprule
        \textbf{Constraint Type} & \textbf{LLaMA-3.3-70B} & \textbf{Kimi-K2-0905} & \textbf{Qwen3-32B} & \textbf{GPT-5} & \textbf{Gemini-2.5-pro} & \textbf{GPT-4.1} & \textbf{Average} & \textbf{Avg Drop} \\
        \midrule
        capital word frequency & 80.8 $\rightarrow$ 45.0 & 74.8 $\rightarrow$ 40.0 & 71.6 $\rightarrow$ 40.0 & 88.0 $\rightarrow$ 60.0 & 80.0 $\rightarrow$ 45.0 & 76.0 $\rightarrow$ 45.0 & 78.5 $\rightarrow$ 45.8 & 32.7 \\
        \textbf{english capital} & \textbf{82.4 $\rightarrow$ 32.0} & \textbf{82.0 $\rightarrow$ 32.0} & \textbf{69.2 $\rightarrow$ 12.0} & \textbf{80.4 $\rightarrow$ 44.0} & \textbf{85.6 $\rightarrow$ 60.0} & \textbf{78.8 $\rightarrow$ 52.0} & \textbf{79.7 $\rightarrow$ 38.7} & \textbf{41.0} \\
        english lowercase & 95.3 $\rightarrow$ 84.2 & 91.6 $\rightarrow$ 68.4 & 90.0 $\rightarrow$ 65.8 & 91.8 $\rightarrow$ 76.3 & 94.5 $\rightarrow$ 76.3 & 94.7 $\rightarrow$ 86.8 & 93.0 $\rightarrow$ 76.3 & 16.7 \\
        repeat prompt & 91.2 $\rightarrow$ 75.0 & 95.0 $\rightarrow$ 70.0 & 92.2 $\rightarrow$ 72.5 & 93.5 $\rightarrow$ 67.5 & 96.8 $\rightarrow$ 82.5 & 93.2 $\rightarrow$ 77.5 & 93.6 $\rightarrow$ 74.2 & 19.4 \\
        two responses & 99.2 $\rightarrow$ 91.7 & 97.1 $\rightarrow$ 87.5 & 92.1 $\rightarrow$ 79.2 & 100.0 $\rightarrow$ 100.0 & 99.2 $\rightarrow$ 91.7 & 97.9 $\rightarrow$ 87.5 & 97.6 $\rightarrow$ 89.6 & 8.0 \\
        number placeholders & 99.6 $\rightarrow$ 96.3 & 98.5 $\rightarrow$ 88.9 & 98.9 $\rightarrow$ 92.6 & 99.6 $\rightarrow$ 96.3 & 98.9 $\rightarrow$ 88.9 & 99.6 $\rightarrow$ 96.3 & 99.2 $\rightarrow$ 93.2 & 6.0 \\
        postscript & 100.0 $\rightarrow$ 100.0 & 98.8 $\rightarrow$ 88.0 & 96.8 $\rightarrow$ 88.0 & 99.2 $\rightarrow$ 92.0 & 100.0 $\rightarrow$ 100.0 & 98.4 $\rightarrow$ 96.0 & 98.9 $\rightarrow$ 94.0 & 4.9 \\
        constrained response & 96.0 $\rightarrow$ 80.0 & 96.0 $\rightarrow$ 80.0 & 100.0 $\rightarrow$ 100.0 & 99.0 $\rightarrow$ 90.0 & 97.0 $\rightarrow$ 90.0 & 96.0 $\rightarrow$ 80.0 & 97.3 $\rightarrow$ 86.7 & 10.6 \\
        json format & 93.5 $\rightarrow$ 58.8 & 89.4 $\rightarrow$ 52.9 & 92.4 $\rightarrow$ 58.8 & 88.8 $\rightarrow$ 52.9 & 95.3 $\rightarrow$ 64.7 & 95.3 $\rightarrow$ 52.9 & 92.5 $\rightarrow$ 56.8 & 35.7 \\
        multiple sections & 93.6 $\rightarrow$ 85.7 & 99.3 $\rightarrow$ 92.9 & 99.3 $\rightarrow$ 92.9 & 100.0 $\rightarrow$ 100.0 & 99.3 $\rightarrow$ 92.9 & 99.3 $\rightarrow$ 92.9 & 98.5 $\rightarrow$ 92.9 & 5.6 \\
        number bullet lists & 100.0 $\rightarrow$ 100.0 & 94.2 $\rightarrow$ 83.9 & 88.7 $\rightarrow$ 77.4 & 99.4 $\rightarrow$ 93.5 & 97.7 $\rightarrow$ 93.5 & 89.0 $\rightarrow$ 80.6 & 94.8 $\rightarrow$ 88.1 & 6.9 \\
        number highlighted sections & 98.8 $\rightarrow$ 87.5 & 96.2 $\rightarrow$ 81.2 & 97.7 $\rightarrow$ 81.2 & 97.3 $\rightarrow$ 81.2 & 92.7 $\rightarrow$ 64.6 & 97.7 $\rightarrow$ 79.2 & 96.7 $\rightarrow$ 79.1 & 17.6 \\
        \textbf{title} & \textbf{100.0 $\rightarrow$ 100.0} & \textbf{98.9 $\rightarrow$ 94.6} & \textbf{98.4 $\rightarrow$ 89.2} & \textbf{99.5 $\rightarrow$ 97.3} & \textbf{98.1 $\rightarrow$ 97.3} & \textbf{99.5 $\rightarrow$ 97.3} & \textbf{99.1 $\rightarrow$ 96.0} & \textbf{3.1} \\
        existence & 97.2 $\rightarrow$ 89.7 & 98.5 $\rightarrow$ 87.2 & 96.4 $\rightarrow$ 79.5 & 99.7 $\rightarrow$ 97.4 & 100.0 $\rightarrow$ 100.0 & 96.2 $\rightarrow$ 84.6 & 98.0 $\rightarrow$ 89.7 & 8.3 \\
        forbidden words & 95.4 $\rightarrow$ 77.1 & 94.4 $\rightarrow$ 66.7 & 86.2 $\rightarrow$ 52.1 & 97.3 $\rightarrow$ 75.0 & 96.5 $\rightarrow$ 72.9 & 96.0 $\rightarrow$ 75.0 & 94.3 $\rightarrow$ 69.8 & 24.5 \\
        keywords frequency & 94.0 $\rightarrow$ 71.8 & 92.1 $\rightarrow$ 61.5 & 84.8 $\rightarrow$ 51.3 & 97.9 $\rightarrow$ 84.6 & 96.0 $\rightarrow$ 71.8 & 93.6 $\rightarrow$ 66.7 & 93.1 $\rightarrow$ 68.0 & 25.1 \\
        letter frequency & 69.7 $\rightarrow$ 45.5 & 70.6 $\rightarrow$ 39.4 & 62.1 $\rightarrow$ 39.4 & 97.0 $\rightarrow$ 87.9 & 75.8 $\rightarrow$ 54.5 & 70.3 $\rightarrow$ 30.3 & 74.2 $\rightarrow$ 49.5 & 24.7 \\
        response language & 97.7 $\rightarrow$ 80.0 & 98.3 $\rightarrow$ 83.3 & 95.7 $\rightarrow$ 76.7 & 98.3 $\rightarrow$ 83.3 & 98.7 $\rightarrow$ 86.7 & 98.3 $\rightarrow$ 83.3 & 97.8 $\rightarrow$ 82.2 & 15.6 \\
        nth paragraph first word & 84.2 $\rightarrow$ 50.0 & 81.7 $\rightarrow$ 50.0 & 80.8 $\rightarrow$ 41.7 & 96.7 $\rightarrow$ 66.7 & 74.2 $\rightarrow$ 41.7 & 77.5 $\rightarrow$ 58.3 & 82.5 $\rightarrow$ 51.4 & 31.1 \\
        number paragraphs & 98.1 $\rightarrow$ 85.2 & 95.2 $\rightarrow$ 81.5 & 88.9 $\rightarrow$ 63.0 & 95.2 $\rightarrow$ 81.5 & 98.1 $\rightarrow$ 92.6 & 88.5 $\rightarrow$ 63.0 & 94.0 $\rightarrow$ 77.8 & 16.2 \\
        number sentences & 87.3 $\rightarrow$ 63.0 & 83.7 $\rightarrow$ 47.8 & 82.5 $\rightarrow$ 37.0 & 97.7 $\rightarrow$ 84.8 & 90.6 $\rightarrow$ 60.9 & 75.6 $\rightarrow$ 39.1 & 86.2 $\rightarrow$ 55.4 & 30.8 \\
        \textbf{number words} & \textbf{91.2 $\rightarrow$ 70.0} & \textbf{87.7 $\rightarrow$ 54.0} & \textbf{83.1 $\rightarrow$ 40.0} & \textbf{96.0 $\rightarrow$ 88.0} & \textbf{95.2 $\rightarrow$ 80.0} & \textbf{85.2 $\rightarrow$ 48.0} & \textbf{89.7 $\rightarrow$ 63.3} & \textbf{26.4} \\
        no comma & 99.1 $\rightarrow$ 92.9 & 98.9 $\rightarrow$ 89.3 & 93.8 $\rightarrow$ 67.9 & 99.3 $\rightarrow$ 92.9 & 99.5 $\rightarrow$ 94.6 & 96.8 $\rightarrow$ 82.1 & 97.9 $\rightarrow$ 86.6 & 11.3 \\
        end checker & 91.9 $\rightarrow$ 69.2 & 92.7 $\rightarrow$ 73.1 & 88.5 $\rightarrow$ 57.7 & 98.8 $\rightarrow$ 88.5 & 91.2 $\rightarrow$ 80.8 & 96.9 $\rightarrow$ 76.9 & 93.3 $\rightarrow$ 74.4 & 18.9 \\
        quotation & 94.9 $\rightarrow$ 60.0 & 97.8 $\rightarrow$ 85.0 & 96.6 $\rightarrow$ 80.0 & 98.0 $\rightarrow$ 85.0 & 98.3 $\rightarrow$ 85.0 & 97.3 $\rightarrow$ 82.5 & 97.1 $\rightarrow$ 79.6 & 17.5 \\
        \bottomrule
    \end{tabular}
    }
\end{table*}

\subsection{Pilot Experiments on \llm{Qwen3-32B}}

We extend our pilot experiments to \llm{Qwen3-32B}.
As shown in~\Cref{fig:poc-results-qwen3-32b}, the instability persists in the larger-scale LLM.
Overall, for the more challenging \textit{Around} relation, the LLM is far less reliable, consistent with the findings of~\citet{zhang2025lifebench-evaluating-length-instruction}.
The instability reveals the vulnerability of LLMs in nuance-oriented reliability, motivating us to study this critical yet underexplored dimension.

\subsection{Category-Wise Reliability}
\label{appx:category-wise-reliability}
To pinpoint the sources of instability, we conduct a fine-grained analysis of the performance degradation under different constraint categories.
\Cref{tab:category_wise_performance} details the drop from standard Accuracy to \reliableat{10} for varying constraint types across representative models. 
The results indicate that the reliability degradation is not entirely uniform, which leads to three key observations:

First, the \textit{Title} constraint proves to be the most manageable. Globally, LLMs show the highest reliability under this constraint, with only a modest drop in performance (averaging \textbf{3.1\%}) for most models. Similar stability is observed in other formatting constraints, such as \textit{Postscript} and \textit{Two Responses}, indicating that simple formatting constraints are generally easy for LLMs to handle.

Second, the \textit{English Capital} constraint (requiring all responses to be in uppercase) is the most difficult.
LLMs show high sensitivity to this constraint, resulting in the largest average drop of \textbf{41.0\%}. 
Models like \llm{Qwen3-32B} deteriorate dramatically (from 69.2\% accuracy to only 12.0\% reliability).
We hypothesize that LLMs may not be well-trained to recognize which tokens contain lowercase letters.

Third, models show substantial differences in reliability under the \textit{Number Words} constraint. Strong models, such as \llm{GPT-5}, maintain relatively stable performance (96.0\% → 88.0\%), whereas others, including \llm{Qwen3-32B}, experience pronounced declines (83.1\% → 40.0\%).
We hypothesize that this is due to tokenization issues, where the counted number of words does not always correspond to the number of tokens used internally by the LLM, potentially hindering its understanding.


\begin{table*}[t]
    \centering
    \small
    \caption{\textbf{The List of 25 Code-Verifiable Constraints.}}
    \label{tab:ifeval-constraints-type-part1}
    \resizebox{0.97\textwidth}{!}{
    \begin{tabular}{>{\centering\arraybackslash}m{1.5cm}
    >{\centering\arraybackslash}m{2cm}
    >{\centering\arraybackslash}m{2cm}
    >{\raggedright\arraybackslash}m{10cm}}
        \toprule
        \textbf{Instruction Group} & \textbf{Instruction} & \textbf{Placeholders} &
        \textbf{Example} \\
        \midrule
        Keywords & Include Keywords & keywords & Write a blog post about how to train a dog that is geared towards kids. Include the keywords ``\{finale\}'' and ``\{less\'' in the post.\\
        \midrule
        Keywords & Keyword Frequency & relation, keyword, frequency & Write a fairy tale about a princess and a dragon, making sure the word ``\{replied\}'' appears \{at least\} \{twice\}.\\
        \midrule
        Keywords & Forbidden Words & forbidden\_words & Can you write a rap that doesn't include the keywords ``\{Yo\}'', ``\{check\}'', and ``\{peace\}''?\\
        \midrule
        Keywords & Letter Frequency & let\_relation, letter, let\_frequency & Make a tweet for playboy's twitter account without using capital letters. Include \{at least\} \{4\} hashtags, starting with ``\{\#\}''\\
        \midrule
        Language & Response Language & language & what is the difference between a levee and an embankment? Please respond to me only in \{Korean\}.\\
        \midrule
        Length Constraints & Number Paragraphs & num\_paragraphs & Write a very angry letter to someone who's been trying to convince you that 1+1=3. There should be exactly \{4\} paragraphs. Separate the paragraphs with \*\*\*.\\
        \midrule
        Length Constraints & Number Words & relation, num\_words & Write a blog post about the best way to get a good night's sleep with \{at least\} \{400\} words.\\
        \midrule
        Length Constraints & Number Sentences & relation, num\_sentences & Write a template with \{less than\} \{7\} sentences for how to calculate the offset of an element in an array.\\
        \midrule
        Length Constraints & Number Paragraphs + First Word in i-th Paragraph & first\_word, num\_paragraphs, nth\_paragraph & How are you doing today? Could you write me exactly 4 paragraphs each separated by two new lines? Please start the first paragraph with the word ``firm''. \\
        \midrule
        Detectable Content & Postscript & postscript\_marker & Write me a template for a product description in the form of a poem and end it with a postscript starting with \{P.P.S\}\\
        \midrule
        Detectable Content & Number Placeholder & num\_placeholders & Draft a blog post about Ripley's Aquarium. Make sure your blog post contains at least \{7\} placeholders represented by square brackets, such as [location].\\
        \midrule
        Detectable Format & Number Bullets & num\_bullets & Make a rubric for a home theater installation targeting moms. Your answer must contain exactly \{4\} bullet points. Use markdown bullet points such as:
        
        * This is point 1\\  
        \midrule
        Detectable Format & Title & -- & Please write a riddle about the inverse function with a title wrapped in double angular brackets, i.e. \textless\textless title\textgreater\textgreater.\\
        \midrule
        Detectable Format & Multiple Sections & section\_spliter, num\_sections & Write a \{4\} section resume for professional clown Phil Larkin. Each section should be explicitly noted as \{Section\} X.\\
        \midrule
        Detectable Format & Choose From & -- & Would you consider yourself to be smart? Choose from:
        
        My answer is yes.
        
        My answer is no.
        
        My answer is maybe.
        
        Just choose one phrase from above as your answer.\\
        
        \midrule
        Detectable Format & JSON Format & -- & What are the advantages and disadvantages of having supernatural powers? Make it short. Wrap the entire output in JSON format. You can use markdown ticks such as \textasciigrave\textasciigrave\textasciigrave.\\
        \midrule
        Combination & Repeat Prompt & prompt\_to\_repeat & \{What is the name of the actor who played Gandalf in Lord of the Rings?\}
        
        First repeat the question above without change of words, then give your answer.\\
        \midrule
        Combination & Two Responses & -- & What is a name that people call God? Please give exactly two different responses. Separate the responses with 6 asterisk symbols: \textasteriskcentered\textasteriskcentered\textasteriskcentered\textasteriskcentered\textasteriskcentered\textasteriskcentered.\\
        \midrule
        Change Cases & All Uppercase & -- & Create a riddle about the name Sheldon using only 10 words. Make sure to only use capital letters in your entire response.\\
        \midrule
        Change Cases & All Lowercase & -- & What is another word for Engravings? Answer in lowercase letters only, throughout your entire answer.\\
        \midrule
        Change Cases & Frequency of All capital Words & capital\_relation, capital\_frequency & Write a serious riddle about trips and stitches in a poem style that includes at least 15 words in all capital letters.\\
        \midrule
        Start with / End with & End Checker & end\_phrase & Write a poem about two people who meet in a coffee shop and end your entire response with the exact phrase ``\{Is there anything else I can help with?\}''\\
        \midrule
        Start with / End with & Quotation & -- & Write a review of IBM's 1956 chess program. Make sure your entire response is wrapped in double quotation marks.\\
        \midrule
        Punctuation & No Commas & -- & Write a limerick about writing a limerick. Don't use any commas in your entire reply.\\
        \bottomrule
    \end{tabular}
    }
\end{table*}



\begin{table*}[t]
\centering
\small
\caption{\textbf{Examples of Revised Infeasible Test Cases from the \textsc{IFEval} Benchmark}. We identify four major categories of issues: (1) flawed implementation of constraint checkers, (2) flawed implementation of constraint arguments, (3) typos in prompts, and (4) unachievable multi-constraint cases.}
\label{tab:revision-examples}
\resizebox{\textwidth}{!}{
\begin{tabular}{p{0.15\linewidth} p{0.23\linewidth} p{0.23\linewidth} p{0.23\linewidth} p{0.2\linewidth}}
\toprule
\textbf{Flaw Type} & \textbf{Original Prompt } & \textbf{Revised Prompt} & \textbf{Reason for Revision} &\textbf{Involved Cases} \\
\midrule
Prompt–checker misalignment
& Write a 2 paragraph critique of the following sentence in all capital letters, no lowercase letters allowed: ``If the law is bad, you should not follow it''. Label each paragraph with PARAGRAPH X.
& Write a 2 paragraph critique \hl{in English} of the following sentence in all capital letters, no lowercase letters allowed: ``If the law is bad, you should not follow i''. Label each paragraph with PARAGRAPH X.
&  For the \textbf{english\_capital} and \textbf{english\_lowercase} constraint, the original test cases did not explicitly specify the use of English. We revised the prompts to align with the checker's evaluation criteria.
& 1021, 1051, 1122, 1132, 1153, 1287, 1516, 1535, 1571, 1593, 1645, 1779, 1813, 1999, 202, 2100, 2391, 2531, 2563, 2943, 296, 3069, 3380, 3434, 3703, 3617, 3456, 3186, 3276
\\
\midrule
Prompt–checker misalignment
& Write two jokes about rockets. Do not contain commas in your response. Separate the two jokes with 6 asterisk symbols: ******.
& Write \hl{two different} jokes about rockets. Do not contain commas in your response. Separate the two jokes with 6 asterisk symbols: ******.
& For the \textbf{two\_responses} constraint, prompts were revised to explicitly require distinct responses, aligning with the checker’s evaluation.
& 1107, 1582, 1591, 3287
\\
\midrule
Prompt–checker misalignment
& Explain the difference between a city and a village in a rap style to a kid. The words with all capital letters should appear at least 10 times. Put the response into \hl{at least} 5 sections, separated using 3 asterisks ***.
& Explain the difference between a city and a village in a rap style to a kid. The words with all capital letters should appear at least 10 times. Put the response into \hl{exactly} 5 sections, separated using 3 asterisks ***.
& For the \textbf{number\_paragraphs} constraint, the evaluation requires exactly n paragraphs, whereas the original natural language prompt only specifies “at least n,” creating a mismatch.
& 2180
\\
\midrule
Prompt–checker misalignment
& Tell a joke that has the words thursday and amalgamation in it, but use Swahili language only, no other language is allowed.
& Tell a joke that \hl{has the words thursday and amalgamation} in it, the response should be primarily \hl{in Swahili}, though you may include a few words from other languages if needed.
& Revised to relax the \textbf{language} requirement: the response should be primarily in the specific language, allowing a few words from other languages if necessary (e.g., English keywords). This ensures the prompt is feasible while preserving the intended linguistic constraint.
& 2309, 3063, 3311
\\
\midrule
Prompt–arguments contradiction
& ... Use the letter o as a keyword in the syntax of the template. The letter o should appear \hl{at least 6 times.}...
& ...Use the letter o as a keyword in the syntax of the template. The letter o should appear \hl{less than 6 times.} ...
(Evaluation augments: \{"let\_relation": "less than", "letter": "o", "let\_frequency": 6\})
& Revised to resolve \textbf{contradictions} between the natural language prompt and evaluation augments.
& 1174, 1217, 1964, 3369, 337
\\
\midrule
Prompt typos
& Create a table with a 7 day trip itinerary for India, and a 7 day trip itinerary for China. Separate them with \hl{exactly 6 asterisks symbols: *******}
& Create a table with a 7 day trip itinerary for India, and a 7 day trip itinerary for China. Separate them with exactly 6 asterisks symbols: ******
& Revised to correct \textbf{typos} in the natural language prompt.
& 1332
\\
\midrule
Incompatible constraints
& \parbox[t]{\linewidth}{%
\textbf{Write a plot for a story about two people who swap fingerprints. Include a title wrapped in double angular brackets\hl{,} i.e. <<title>>. In your response please \hl{avoid using commas}.}\\
First, repeat the request above word for word without change.\\
Do not say any words or characters before repeating the request above.\\
After you repeated the request, you can give your response next.%
}
& \parbox[t]{\linewidth}{\textbf{Write a plot for a story about two people who swap fingerprints. Include a title wrapped in double angular brackets, i.e. <<title>>.}\\
First, repeat the request above word for word without change.\\
Do not say any words or characters before repeating the request above.\\
After you repeated the request, you can give your response next.}
& The original prompt imposes \textbf{multiple constraints that are mutually incompatible}, making it unachievable. For instance, it requires repeating a specific text fragment containing commas while simultaneously prohibiting the use of commas in the response.
& 1627, 2216, 3305, 3371, 3718, 374
\\
\bottomrule
\end{tabular}
}
\end{table*}


\begin{table*}[!t]
    \centering
    \small
    \setlength{\tabcolsep}{3pt}
    \renewcommand{\arraystretch}{1}
    \caption{\textbf{Model Details}.}
    \label{tab:model-details}
    \resizebox{1\textwidth}{!}{
    \begin{tabular}{lccccc}
    \toprule
    \textbf{Model}  &  \textbf{Vendor} &  \textbf{Release Date} & \textbf{\# Params}  &  \textbf{Context Size} & \textbf{Knowledge Cutoff} \\
         \mymidrule{1-6} \rowcolor[gray]{0.9}
    \multicolumn{6}{c}{\textbf{\textit{Proprietary LLMs (API-Based)}}}\\ 
    \mymidrule{1-6}
    GPT-5               & OpenAI & 2025-08 & ? & 400k & 2024-09 \\
    GPT-5-mini          & OpenAI & 2025-08 & ? & 400k & 2024-05 \\
    GPT-5-nano          & OpenAI & 2025-08 & ? & 400k & 2024-09 \\
    GPT-5-chat-latest   & OpenAI & 2025-08 & ? & 125K & 2024-09  \\
    Gemini-2.5-Pro      & Google & 2025-06 & ? & 1M & 2025-01 \\
    Gemini-2.5-Flash    & Google & 2025-06 & ? & 1M & 2025-01 \\
    Gemini-2.5-Flash-Lite & Google & 2025-06 & ? & 1M & 2025-01 \\
    o4-mini             & OpenAI & 2025-04 & ? & ? & 2024-06 \\
    GPT-4.1             & OpenAI & 2025-04 & ? & 1M & 2024-06 \\
    GPT-4.1-mini        & OpenAI & 2025-04 & ? & 1M & 2024-06 \\
    GPT-4.1-nano        & OpenAI & 2025-04 & ? & 1M & 2024-06 \\
    o3                  & OpenAI & 2025-04 & ? & 195K & 2024-06 \\
    o3-mini             & OpenAI & 2025-01 & ? & 195K & 2023-10 \\
    GPT-4o-2024-11-20   & OpenAI & 2024-11 & ? & 125K & 2023-10 \\
    GPT-4o              & OpenAI & 2024-11 & ? & 125K & 2023-10 \\
    GPT-40-2024-05-13   & OpenAI & 2024-11 & ? & 125K & 2023-10 \\
    o1-mini             & OpenAI & 2024-09 & ? & 125K & 2023-10 \\
    GPT-4o-mini         & OpenAI & 2024-07 & ? & 125K & 2023-10 \\
    GPT-3.5-turbo-0125  & OpenAI & 2024-01 & ? & 16K  & 2021-09 \\
    GPT-3.5-turbo-1106  & OpenAI & 2023-11 & ? & 16K  & 2021-09 \\
    \mymidrule{1-6} \rowcolor[gray]{0.9}
    \multicolumn{6}{c}{\textbf{\textit{Open-Source LLMs}}}\\  
    \mymidrule{1-6}
    Qwen3-Next-80B-A3B-TK       & Alibaba   & 2025-09 & 80B (A3B)   & 262K & ? \\
    Kimi-K2-0905                & Moonshot  & 2025-09 & 1T(A32B)    & 256K & 2025-3 \\
    Qwen3-Next-80B-A3B-Instruct & Alibaba   & 2025-09 & 80B (A3B)   & 256K & ?\\
    GPT-oss-120b                & OpenAI    & 2025-08 & 120B (A5.1B)& 128K & 2024-06 \\
    GPT-oss-20b                 & OpenAI    & 2025-08 & 20B (A3.6B) & 128K & 2024-06 \\
    GLM-4.5                     & Z.AI      & 2025-07 & 355B (A32B) & 128K & ? \\
    Qwen3-30B-A3B               & Alibaba   & 2025-07 & 30B (A3B)   & 256K & ? \\
    GLM-4.5-Air                 & Z.AI      & 2025-07 & 106B        & 128K & ?\\
    Qwen3-235B-A22B             & Alibaba   & 2025-07 & 235B (A22B) & 256K & ? \\
    Qwen3-14B                   & Alibaba   & 2025-05 & 14.8B       & 32K  & ? \\
    Qwen3-8B                    & Alibaba   & 2025-05 & 8B          & 128K & ? \\
    Qwen3-32B                   & Alibaba   & 2025-05 & 32B         & 128K & ? \\
    Qwen3-4B                    & Alibaba   & 2025-04 & 4B          & 32K  & ? \\
    Qwen3-1.7B                  & Alibaba   & 2025-04 & 2B          & 32k  & ?\\
    Qwen3-0.6B                  & Alibaba   & 2025-04 & 0.8B        & 32k  & ?\\
    Gemma-3-IT-27B              & Google    & 2025-03 & 27B         & 128K & 2024-08 \\
    Gemma-3-IT-12B              & Google    & 2025-03 & 12B         & 128K & 2024-08 \\
    DeepSeek-V3.1               & DeepSeek  & 2025-01 & 671B (A37B) & 128K & 2025-07 \\
    DeepSeek-R1                 & DeepSeek  & 2025-01 & 671B (A37B) & 128K & 2024-12 \\
    LLaMA-3.3-70B-IT            & Meta      & 2024-12 & 70B         & 128K & 2023-12 \\
    Qwen2.5-72B-Instruct        & Alibaba   & 2024-09 & 72B         & 131k & ?\\
    Qwen2.5-32B-Instruct        & Alibaba   & 2024-09 & 32B         & 128K & ? \\
    Qwen2.5-14B-Instruct        & Alibaba   & 2024-09 & 14.7B       & 128K & ?\\
    Qwen2.5-7B-Instruct         & Alibaba   & 2024-09 & 7B          & 32K  & ? \\
    LLaMA-3.1-70B-Instruct      & Meta      & 2024-07 & 70B         & 128K & 2023-12\\
    LLaMA-3.1-8B-Instruct       & Meta      & 2024-07 & 8B          & 128K & 2023-12 \\
    \bottomrule
    \end{tabular}
    }
\end{table*}

\clearpage

\begin{table*}
    \centering
    \caption{\textbf{System Prompts Used in~\Cref{appx:system-prompt-impact}}.}
    \label{tab:system-prompt-settings}
    \begin{tabular}{p{0.3\textwidth}p{0.3\textwidth}p{0.3\textwidth}}
    \toprule
    \textbf{Persona 1}     & \textbf{Persona 3} & \textbf{Persona 2} \\
    \midrule
    \parbox[t]{\linewidth}{
You are a helpful assistant serving the following persona, and you should adjust your response to the persona.\\
A person who is interested in European Union law, particularly the concept of EU citizenship and its rights. \\
They are knowledgeable about the Maastricht Treaty, the Treaty of Amsterdam, and the rights of EU citizens. \\
They are also interested in the concept of 'union citizenship' and how it complements national citizenship. \\
They may have a background in law or politics and are interested in understanding how EU laws are implemented and enforced.
}    & \parbox[t]{\linewidth}{
You are a helpful assistant serving the following persona, and you should adjust your response to the persona.\\
A finance manager who is interested in understanding the different inventory accounting methods used in business, particularly the advantages and disadvantages of each method.\\
They are looking for a detailed explanation of the cost of goods sold and how it affects inventory physical flow and value.\\
They are also interested in learning about the specific examples given in the text and how they can apply these concepts in their own business.
} & \multirow{3}{*}{
\parbox[t]{\linewidth}{
You are a helpful assistant serving the following persona, and you should adjust your response to the persona.\\
A conservation biologist who specializes in the study of aquatic ecosystems, specifically focusing on the restoration of migratory fish habitats and the monitoring of aquatic species.\\
They have extensive experience working with the Virginia Department of Game and Inland Fisheries (DGIF) and have a deep understanding of the complexities of dam removal and its impact on aquatic habitats.\\
They have a passion for conservation and have dedicated their career to protecting and restoring natural ecosystems.\\
They have a keen eye for detail and are skilled at analyzing data to identify patterns and trends in the ecosystem.\\
They have a strong work ethic and are committed to making a positive impact on the environment.
}

} \\
\cmidrule(lr){1-2}
\textbf{Python} & \textbf{Following-Instructions} \\
\cmidrule(lr){1-2}
\parbox[t]{\linewidth}{
You are a highly skilled and experienced software engineer specializing in Python programming.\\ \\ \\
}  & You are a highly skilled and experienced assistant in following the user's instructions. & \\
\bottomrule
    \end{tabular}

\end{table*}


\begin{table}[t]
    \centering
    \caption{\textbf{Composition of the Dataset Used in~\Cref{subsec:training-based} for the \textit{Cousin Prompts} Case}.}
    \label{tab:train-dataset-composition}
    \resizebox{0.42\textwidth}{!}{
    \begin{tabular}{M{0.4\columnwidth} M{0.3\columnwidth}}
        \toprule
        \textbf{Augmentation} & \textbf{Count} \\
        \midrule
        Rephrasing   & 2,303 \\
        Distractor Addition    & 2,274 \\
        C/T Reconfig  & 2,046 \\
        \bottomrule
    \end{tabular}
    }
\end{table}

\section{Training-Based Methods}
\label{appx:training-methods}

\subsection{Training Details}
\label{appx:training-details}

In this section, we detail the training recipe for the experiments in~\Cref{subsec:training-based}.
Recall that we resort to supervised fine-tuning as the training method, which teaches the LLMs to mirror the expected responses.
This is supervised by a cross-entropy loss on the response tokens.
Formally, let $x$ denote the input prompt (after applying the chat template) and $y = (y_1, \dots, y_T)$ the corresponding target response consisting of $T$ tokens.
The model parameterized by $\theta$ defines a conditional distribution $p_\theta(y_t \mid x, y_{<t})$ over the next token given the input and previously generated tokens.
The supervised fine-tuning objective minimizes the negative log-likelihood of the ground-truth responses, \ie,
\begin{equation}
\mathcal{L}_{\mathrm{SFT}}(\theta)
= - \mathbb{E}{(x, y) \sim \mathcal{D}} \sum_{t=1}^{T} \log p_\theta(y_t \mid x, y_{<t}),
\end{equation}
where $\mathcal{D}$ denotes the training dataset.
This objective is equivalent to minimizing the token-level cross-entropy loss between the predicted distribution and the true response sequence.
During optimization, we employ teacher forcing, feeding the gold tokens $y_{<t}$ at each step, to stabilize training and accelerate convergence.

We train the \llm{Qwen2.5-7B-Instruct} using the Swift framework~\citep{zhao2025swift}.
We use LoRA~\citep{hu2022lora} for parameter-efficient fine-tuning.
We introduce adapters, with rank $r=16$ and coefficient $\alpha=32$, to all linear modules.
Model parameters are updated using the AdamW optimizer~\citep{loshchilov2017adamw}.
The peak learning rate is 0.00005, and the global batch size is 64.
To ensure training stability, we set a warmup ratio of 0.05, cosine learning rate decay, and gradient clipping of 1.0.
We save checkpoints every 20 steps throughout the training process.

We include two dataset settings in the SFT experiments:
(1) general-domain instruction-following, for which we use the representative alpaca~\citep{alpaca} as the data source;
(2) self-curated datasets tailored for improving reliability, for which we employ another pool of cousin prompts for the \dataset{IFEval}.
Across the two settings, the system prompt is consistently set to the default one for both training and evaluation.

\begin{figure}[h]
    \centering
    \includegraphics[width=0.47\textwidth]{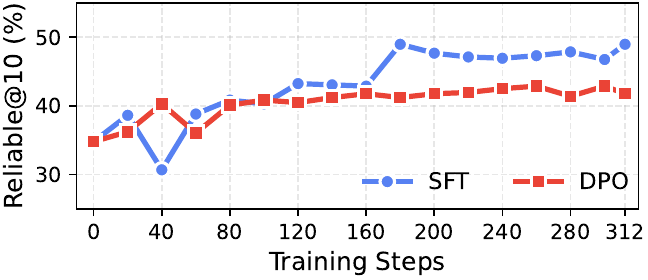}
    \vspace{-0.5em}
    \caption{\textbf{Training Methods Affects Reliability Performance on \ourbench in Varying Fashion}.}
    \label{fig:training-effects-dpo-vs-sft}
    \vspace{-1em}
\end{figure}

We further elaborate on the curation process for the \textit{Cousin Prompts} training set, following the same augmentation-then-filtering paradigm introduced in~\Cref{sec:dataset-curation}. 
For the rephrasing augmentation, we reuse the 16 replicates of cousin prompts described in~\Cref{appx:relatk-scalability}. 
For the other two augmentation methods, each original prompt is expanded into five additional replicates. 
After augmentation, we first remove any prompts that exactly match those in~\ourbench, and then apply the automated validity check. 
To maintain balance across the three augmentation methods, we randomly down-sample the cousin prompts belonging to each augmentation method to 2,400 examples.

We then employ the \llm{LLaMA-3.3-70B-Instruct} as the teacher model, and run sampling 16 times with a temperature of 1.0.
Then, we account for only those prompts with at least one response that can pass the evaluation.
The resulting dataset composition is presented in~\Cref{tab:train-dataset-composition}.
We run SFT on this dataset for three epochs, totaling 312 training steps.
For the \textit{Alpaca} dataset, we apply the same configuration, training on a randomly sampled subset of the complete dataset.
This setup enables us to isolate the effect of the training data itself, independent of other confounding factors, on reliability.

\subsection{Training Methods: SFT vs. DPO}
\label{appx:vs-dpo}

Additionally, we advance an exploratory attempt around distinct training methods.
We include the direct preference optimization (DPO) algorithm~\citep{rafailov2024dpo}, an alignment method that directly optimizes a model's parameters toward preferred responses without requiring explicit reward modeling.
Unlike reinforcement learning from human feedback (RLHF)~\citep{ouyang2022instructgpt}, which relies on a separately trained reward model and policy optimization, DPO leverages pairwise preference data.
This encourages the model to assign higher relative likelihood to preferred responses compared to rejected ones, enabling alignment without explicit reward modeling.
The experiments are conducted on the same \textit{Cousin Prompts} dataset.
For DPO, the rejected samples are collected from \llm{LLaMA-3.3-70B-Instruct}. 
When multiple rejected responses are available, we randomly select from those that fail to meet the evaluation criteria. 
If none explicitly fail, we instead choose the response with the largest character count, as it intuitively represents the least efficient way of satisfying the requirements.
All other training configurations are shared across the two training methods, including the learning rate.

As illustrated in~\Cref{fig:training-effects-dpo-vs-sft}, DPO in this setting does not further boost the reliability and even underperforms compared to vanilla SFT.
We hypothesize that this phenomenon arises from the construction of the rejected responses. 
Specifically, the current setup may fail to capture the nuanced discrepancy that the policy model should learn to minimize between preferred and rejected outputs.
In the context of instruction-following, a rejected response may still reflect partial progress toward fulfilling the given constraints. 
Penalizing such outputs too strongly could inadvertently steer the model away from learning the correct alignment direction.

We leave this non-trivial exploration of training methods in the Appendix, as we cannot draw conclusions about best practices.
Future works may extend this investigation to other training paradigms, \eg, reinforcement learning from verifiable rewards~\citep{lambert2024tulu-3-rlvr,guo2025deepseek-r1,zou2025eifbench}.

\clearpage

\begin{promptboxc}[prompt:task-alteration]{
Task Reconfiguration
}{\fontsize{10}{12}\selectfont
You are a helpful assistant that revise the prompt to test an LLMs' generalization ability.\\
\\
The prompt typically contain one task and a few constraints. The only part that should be revised is the task description.\\
\\
You are free to revise the task description but keep the new task compatible with the original constraints. Typically, the revised task should be more challenging than the original task.\\
\\
You should make the minimal changes to the constraints unless absolutely necessary, such that the new prompt is still able to request the same set of constraints.\\
\\
Both the responses to the original prompt and the new prompt will be evaluated by the following function:\\
\textcolor{blue}{\{evaluation\_function\_implementation\}}\\
\\
And the necessary arguments for the evaluation function (i.e., the requirements to be satisfied by the response) are:\\
\textcolor{blue}{\{evaluation\_function\_arguments\}}\\
\\
The original prompt is:\\
\textcolor{blue}{\{prompt\}}\\
\\
You should directly output the new prompt, without any other text.
}
\end{promptboxc}

\begin{promptboxc}[prompt:rephrasing]{
Rephrasing the Prompts
}{\fontsize{10}{12}\selectfont
You are a helpful assistant whose task is to rephrase a given prompt into another prompt.\\
\\
You will be provided with:\\
1. A complete natural language prompt consisting of a task and one or more constraints.\\
2. A set of evaluation functions that will be used to evaluate responses to the prompt.\\
3. The arguments for the evaluation functions.\\
\\
\#\#\# Your task\\
Rewrite the given prompt while following these rules:\\
- Treat the input as a single, complete prompt. Keep all constraints intact, but rephrase them naturally.\\
- Do not alter numbers, thresholds, links, counts, or formatting requirements.\\
- Do not add new constraints or information.\\
- Preserve explicit structural rules (e.g., ``repeat exactly'', ``use this format'', ``separate items with ******'').\\
- Always output the rewritten prompt in English, even if the original is in another language.\\
- If a constraint specifies a casing requirement (e.g., ALL CAPS, all lowercase), rephrase it using normal sentence casing.\\
- For constraints in the **repeat\_prompt** category, do not modify the sentence to be repeated (the one provided in the arguments).\\
\\
---\\
\\
\#\#\# Input\\
**Original prompt:**\\
\textcolor{blue}{\{prompt\}}\\
\\
**Evaluation functions:**\\
\textcolor{blue}{\{evaluation\_function\_implementation\}}\\
\\
**Arguments:**\\
\textcolor{blue}{\{evaluation\_function\_arguments\}}\\
\\
---\\
\\
\#\#\# Output\\
Directly produce the rewritten prompt, without any additional text.
}
\end{promptboxc}

\begin{promptboxc}[prompt:distractor]{
Add a Distractor Constraint
}{\fontsize{10}{12}\selectfont
You are a helpful assistant tasked with adding one additional distractor constraint to the original prompt.\\
\\
\#\#\# Instructions:\\
- Preserve all contents of the original prompt exactly as they are.  \\
- At the end of the new prompt, append one extra constraint sentence.  \\
- This extra constraint must introduce an additional **format requirement** for the response.  \\
- The new constraint must not interfere with or alter the original constraints.  \\
\\
\#\#\# Original Constraints:\\
\textcolor{blue}{\{constraints\}}\\
\\
\#\#\# Evaluation Functions:\\
\textcolor{blue}{\{evaluation\_functions\}}\\
\\
\#\#\# Original Prompt:\\
\textcolor{blue}{\{prompt\}}
}
\end{promptboxc}

\begin{promptboxc}[prompt:constraint-alteration]{
Constraint Reconfiguration
}{\fontsize{10}{12}\selectfont
You are a helpful assistant that revise the prompt to test an LLMs' generalization ability.\\
\\
The prompt typically contain one task and a few constraints. You are allowed to revise the constraints and correspondingly adjust the evaluation arguments. If necessary, you can also revise the task description.\\
\\
If multiple constraints exist, you should revise at least one of them.\\
\\
Typically, the revised constraints should be **more challenging** than the original constraints, but not beyond the **allowable range** of the evaluation functions.\\
\\
Keep using the original constraint types of the original prompt and the original evaluation functions. Only change the fill-in values (arguments) of the constraints.\\
\\
To help you better understand the constraints, here are the evaluation functions that will be used to evaluate the responses to the original prompt and the new prompt:\\
\textcolor{blue}{\{evaluation\_function\_implementations\}}\\
\\
The original prompt is:\\
\textcolor{blue}{\{prompt\}}\\
\\
The original instruction id list is:\\
\textcolor{blue}{\{instruction\_id\_list\}}\\
\\
The evaluation arguments for the original prompt are:\\
\textcolor{blue}{\{evaluation\_function\_arguments\}}\\
\\
You should output a list of JSON objects with the following fields:\\
- "random\_seed": The random seed you choose to generate the new prompt.\\
- "reasoning": The reasoning process for generating the new prompt and the new kwargs.\\
- "revised\_prompt": The new prompt.\\
- "revised\_instruction\_id\_list": The new instruction id list. If you have added or removed some constraints, you should also update the instruction id list. The length of the list should be the same as that of the revised kwargs.\\
- "revised\_kwargs": The new kwargs in string format, which should be valid JSON. The format should be the same as the original kwargs as follows:\\
\textcolor{blue}{\{evaluation\_function\_arguments\_format\}}\\
\\
The total number of the revised prompts should be \textcolor{blue}{\{num\_prompts\}}. The revised prompts should be **distinct** in constraints, **reasonable**, and increasingly difficult.\\
\\
Keep in mind that do not use constraints beyond the provided evaluation functions.
}
\end{promptboxc}

\begin{promptboxc}[prompt:sanity-checker]{
Checking the Integrity of Test Cases
}{\fontsize{10}{12}\selectfont
You are tasked with verifying whether a given test case for evaluating LLMs' instruction-following behavior is valid.\\
\\
Each test case consists of:\\
- **Prompt**: The instruction provided to the LLM.\\
- **Evaluation Arguments**: The specific parameters required by the evaluation function.\\
- **Evaluation Function**: The function that checks the LLM's response against the evaluation arguments and returns a boolean judgment.\\
\\
Your task is to determine if the test case is valid by checking the following conditions:\\
\\
A test case is valid only if it satisfies all three conditions below:\\
\\
1. **Argument Reflection**  \\
\   Every evaluation argument must be explicitly required or constrained in the Prompt.  \\
\   If the Prompt does not impose a constraint corresponding to a given evaluation argument, the test case is invalid.\\
\   You should understand how the evaluation function works and the constraints (evaluation arguments) to be evaluated.\\
\\
2. **Format Compatibility**  \\
\   The Prompt's format and placeholder conventions must allow a response that can pass the evaluation function.  \\
\   - For arguments of type `keywords`, `keyword`, `forbidden\_words`, or `prompt\_to\_repeat`, the Prompt must clearly require or forbid these elements in such a way that the evaluation function can check them.\\
\   - Be careful with the punctuation marks.\\
\\
3. **Logical Consistency**  \\
\   Logically, the context in the Prompt (e.g., the tasks or other constraints that are not evaluated) must not contradict the constraints to be evaluated.\\
\   All constraints are independent of each other and should be met at the same time. Do not interpret one constraint conditioned on another.\\
\   For example, requiring JSON-formatted output but additionally asking for a plain-text line is invalid.\\
\\
\#\#\# Notes:\\
- Your judgement should be harsh and rigorous. You should not be lenient or allow for any exceptions. Default to "is\_valid": false.\\
- A stricter or looser Prompt constraint is **not** equivalent to an Evaluation Argument.\\
- The test case is valid only if **all three conditions** above are satisfied.\\
\\
\#\#\# Output:\\
Return your judgment in the following JSON format:
\begin{verbatim}
```json
{{
    "reasoning": "Your reasoning process",
    "is_valid": true/false
}}
```
\end{verbatim}

---\\
\\
Prompt:\\
\textcolor{blue}{\{prompt\}}\\
\\
Evaluation Arguments:\\
\textcolor{blue}{\{evaluation\_args\}}\\
\\
Evaluation Function:\\
\textcolor{blue}{\{evaluation\_function\}}
}
\end{promptboxc}

\end{document}